\documentclass[preprint,aps,amsmath,nofootinbib]{revtex4-1}
\usepackage{graphicx,array,dcolumn}
\usepackage{calc,tabularx, epsfig,mathrsfs}
\usepackage{hyperref}

\usepackage{amsmath,verbatim,enumerate}
\usepackage{amssymb}
\usepackage{wasysym}
\usepackage{xcolor}
\usepackage{multirow}
\usepackage{xspace}
\usepackage{slashed}
\allowdisplaybreaks[1]
\newlength{\figurewidth}
\newcommand{\beq}{\begin{equation}}
\newcommand{\eeq}{\end{equation}}
\newcommand{\bea}{\begin{eqnarray}}
\newcommand{\eea}{\end{eqnarray}}
\newcommand{\ba}{\begin{array}}
\newcommand{\ea}{\end{array}}

\newcommand{\mn}{{\mu\nu}}

\newcommand{\pt}{\partial}

\newcommand{\al}{\alpha}

\newcommand{\ep}{\epsilon}
\newcommand{\ta}{\theta}
\newcommand{\lam}{\lambda}
\newcommand{\Lam}{\Lambda}

\newcommand{\de}{\delta}
\newcommand{\D}{\Delta}
\newcommand{\OM}{\Omega}

\newcommand{\sg}{\sigma}

\makeatother
\begin{document}
%
% define Title, Author, Address, Preprint#
%%%%%%%%%%%%%%%%%%%%%%%%%%%%%%%%%%%%%%%%%%%%%%%
\title{Lorentzian quantum cosmology in novel Gauss-Bonnet gravity from Picard-Lefschetz methods}
\setlength{\figurewidth}{\columnwidth}
%%%%%%%%%%%%%%%%%%%%%%%%%%%%%%%%%%%%%%%%%%%%%%%
%
\author{Gaurav Narain$\,{}^a$}
\email{gaunarain@gmail.com}
\author{Hai-Qing Zhang$\,{}^{a,b}$}
\email{hqzhang@buaa.edu.cn}

\affiliation{${}^a$ Center for Gravitational Physics, Department of Space Science, 
Beihang University, Beijing 100191, China. \\
${}^b$ International Research Institute for Multidisciplinary Science, Beihang University, 
Beijing 100191, China}

%
%%%%%%%%%%%%%%%%%%%%%%%%%%%%%%%%%%%%%%%%%%%%%%%
\begin{abstract}
In this paper we study some aspects of classical and quantum cosmology 
in the novel-Gauss-Bonnet (nGB) gravity in four space-time dimensions. Starting 
with a generalised Friedmann-Lema\^itre-Robertson-Walker (FLRW) 
metric respecting homogeneity and isotropicity
in arbitrary space-time dimension $D$, we find the action of theory 
in four spacetime dimension where the limit $D\to4$ is smoothly obtained after an 
integration by parts. The peculiar rescaling 
of Gauss-Bonnet coupling by factor of $D-4$ results in a non-trivial contribution 
to the action. We study the system of equation of motion to 
first order nGB coupling. We then go on to compute the transition probability 
from one $3$-geometry to another directly in Lorentzian signature. 
We make use of combination of WKB approximation and 
Picard-Lefschetz (PL) theory to achieve our aim. PL theory 
allows to analyse the path-integral directly in Lorentzian signature 
without doing Wick rotation. Due to complication caused by non-linear nature 
of action, we compute the transition amplitude to first order in nGB coupling.
We find non-trivial correction coming from the nGB coupling to the transition 
amplitude, even if the analysis was done perturbatively. 
We use this result to investigate the case of classical boundary 
conditions. 
\end{abstract}
%%%%%%%%%%%%%%%%%%%%%%%%%%%%%%%%%%%%%%%%%%%%%%%
\maketitle

%%%%%%%%%%%%%%%%%%%%%%%%%%%%%%%%%%%%%%%%%%%%%%%
\section{Introduction}
\label{intro}
%%%%%%%%%%%%%%%%%%%%%%%%%%%%%%%%%%%%%%%%%%%%%%%

General relativity although enjoys the merit of explaining a wide range of 
physical phenomena over a large range of distance, however its validity 
becomes questionable beyond these regimes where it is expected to get modified.
For example at ultra high energies motivated by lack of renormalizabilty 
of GR it is noticed that addition of higher-derivative terms 
\cite{Stelle:1976gc,Salam:1978fd,Julve:1978xn} results in a better 
ultraviolet behavior of resulting quantum theory. It however comes 
with their own bag of issues regarding lack of unitarity. 
Some efforts have been made in \cite{Narain:2011gs,Narain:2012nf,Narain:2017tvp,Narain:2016sgk},
in asymptotic safety approach 
\cite{Codello:2006in,Niedermaier:2009zz}
and `\textit{Agravity}' \cite{Salvio:2014soa}. Such unitarity 
problems arises as the theory has more than two time-derivatives. 
Lovelock gravity \cite{Lovelock:1971yv,Lovelock:1972vz,Lanczos:1938sf} 
are a special class of higher-derivative gravity 
where equation of motion remains second order in time. 

In four spacetime dimension the Lovelock gravity also known 
as Gauss-Bonnet gravity is topological and doesn't 
contribute in the dynamical evolution of metric. 
However, they play a key role in path-integral quantization 
of gravity where it is used to classify topologies. 
Motivated by works of \cite{Mardones:1990qc,Torii:2008ru} it is 
observed that Gauss-Bonnet gravity can contribute non-trivially 
if its coupling is rescaled by factor of $D-4$ (where $D$ is 
spacetime dimensionality) \cite{Glavan:2019inb}. 
Such rescaling introduces non-trivial features 
coming from Gauss-Bonnet in four spacetime dimensions. 
This has generated tremendous interest in novel Gauss-Bonnet gravity.

The novel Gauss-Bonnet gravity \cite{Glavan:2019inb} action is following
\bea
\label{eq:act}
S = \frac{1}{16\pi G} \int {\rm d}^Dx \sqrt{-g}
\biggl[
-2\Lam + R + \frac{\al}{D-4} 
\biggl( R_{\mu\nu\rho\sg} R^{\mu\nu\rho\sg} - 4 R_\mn R^\mn + R^2 \biggr)
\biggr] \, , 
\eea
where $G$ is the Newton's gravitational constant, 
$\Lam$ is the cosmological constant term,  
$\al$ is the Gauss-Bonnet coupling and $D$ is spacetime dimensionality. 
The Gauss-Bonnet coefficient has been defined with a 
$(D-4)$ factor in denominator. The mass dimensions of 
various couplings are: 
$[G] = M^{2-D}$, $[\Lam] = M^2$ and $[\al] = M^{-2}$. 

It is seen that an integration by parts gets rid of $(D-4)$ factors
leaving behind an action with a well-defined $D\to4$ limit
\cite{Lu:2020iav,Hennigar:2020lsl}. Here the authors do a Kaluza-Klein dimensional 
reduction where the manifold is cross-product of two spaces
${\cal M}_D = {\cal M}_4 \bigotimes {\cal M}_{D-4}$, 
thereby implying that the full metric can be written as a 
four-dimensional metric on ${\cal M}_4$ and extra-dimension piece on ${\cal M}_{D-4}$. 
They notice that taking limit $D\to4$ after an integration by parts 
leads a well-defined action which is Horndeski  type gravity. 
A similar study was conducted in \cite{1794944} using ADM decomposition
realises that for a well-defined limit and a consistent theory in four dimensions
one either break (a part of) the diffeomorphism invariance or have an extra degree of freedom 
\cite{1794944}. In doing a KK reduction it is seen that the four dimensional 
action retains a \textit{memory} of the higher-dimensional manifold, 
which shows up as an appearance of additional scalar field. 

Inspired by these studies we decided to explore quantum aspects of 
novel Gauss-Bonnet gravity in a cosmological setting
We start by considering a generic metric respecting 
spatial homogeneity and isotropicity in $D$-spacetime 
dimensions. It is a generalisation of FLRW metric in $D$-dimensions 
consisting of two unknown time-dependent functions: lapse 
and scale-factor. This is mini-superspace approximation of the metric.
On plugging this metric in novel Gauss-Bonnet 
gravity action and performing integration by parts, we are left with a
mini-superspace action of theory where a well-defined 
$D\to4$ limit can be taken \cite{Narain:2020qhh}. This action contains 
non-trivial contribution from the Gauss-Bonnet term.
This process of obtaining well-defined $4$-dimensional action 
doesn't involve KK type dimensional reduction as in \cite{Lu:2020iav,Hennigar:2020lsl}. 
As a result the  
$4$-dimensional action doesn't have an additional scalar-field 
which is like a \textit{memory} of higher-dimensional manifold. 

In this paper we study the quantum gravity path-integral 
to compute the transition amplitude from one $3$-geometry 
to another, and investigate the non-trivial contributions coming from 
the novel-Gauss-Bonnet gravity. Usually to study such transitions one has 
to study the behavior of the following path-integral 
\beq
\label{eq:EucQGAct}
G[g_1, g_2]= \int_{{\cal C}} {\cal D} g_\mn \exp \left(-I[g_\mn] \right) \, .
\eeq
Here $g_\mn$ is the metric whose gravitational action appears 
in the corresponding exponential and is given by $I[g_\mn]$.
This is Euclideanised version of the original Lorentzian path-integral 
where the temporal part of the metric has been Wick rotated 
in order to have a well-defined convergent path-integral 
along the contour ${\cal C}$. In flat spacetime there is a meaningful 
time co-ordinate and enjoys the 
properties of global symmetries to cast Lorentz group in to 
a compact rotation group under a transformation of time co-ordinate. 
This is hard to replicate in generic curved spacetime. 
In a sense Wick-rotation (a process of defining a convergent 
path-integral by transforming the highly oscillatory path-integral in Lorentzian 
signature to euclidean) in quantum field theory (QFT) on flat 
spacetime is more natural to implement than in curved spacetime
where `time' is just a parameter. The $+i\ep$-prescription by 
Feynman in flat spacetime QFT is a systematic way to choose 
a contour in complexified spacetime, which is done in such a manner 
so that contour doesn't cross the poles of the free theory propagator.
This offers relevant convergence to an otherwise highly oscillatory integral 
and naturally implements causality in path-integral 
in a systematic manner by requiring that the 
euclideanised version of two-point function must satisfy Osterwalder-Schrader positivity. 
Such benefits of flat spacetime is hard to replicate in 
generic Lorentzian spacetime, and it gets even more 
involved when spacetime becomes dynamical due to gravity
and/or gravitational field is also quantized. 
A possibility exists as to do 
a Wick rotation sensibly and obtaining the Lorentizian case from Euclidean 
by properly implementing Wick rotation in curved spacetime 
\cite{Candelas:1977tt,Visser:2017atf,Baldazzi:2019kim,Baldazzi:2018mtl}. 
However, this direction is still in its infant stages and more work needs to be done.

Picard-Lefschetz theory offers a way to handle such kind of 
oscillatory path-integrals. In a sense it is a generalization of standard
Wick-rotation where the process is adapted accordingly to deal with 
generic curved spacetime. Here one study them by integrating them along the 
path of steepest descent in the complexified plane 
where the contour is uniquely obtained by using generalised flow 
equation in complex plane. Such steepest descent flow lines are 
termed \textit{Lefschetz thimbles}. Early attempts 
making use of  knowledge of steepest descent contours occurred in the context of 
Euclidean quantum gravity \cite{Hawking:1981gb,Hartle:1983ai}. 

Motivation to study euclideanised gravitational path-integral was an 
expectation that similar to flat spacetime QFT one will have relevant 
convergence. This is a mistake. Gravitational path-integral are 
non-trivial. Apart from dealing with usual issues of path-integral measure, 
gauge-invariance (gauge-fixing), regularization, renormalizability 
and boundary conditions; it is equally important to 
choose a contour of integration carefully for necessary convergence.
This last bit is obscure in curved spacetime, where the standard 
Feynman $+i\ep$-prescription (which works in flat spacetime QFT)
no longer offers reliable results. 

Picard-Lefschetz theory offers a systematic way to find this 
integration contour in a generic spacetime where the 
gravitational path-integral becomes absolutely convergent.
This has been made use of in the simple models of quantum cosmology 
\cite{Feldbrugge:2017kzv,Feldbrugge:2017fcc,Feldbrugge:2017mbc},
where the authors studied path-integral  
in the mini-superspace approximation. Earlier attempts employing 
similar strategy but in euclidean quantum cosmology
goes back to 1980s \cite{Vilenkin:1982de,Vilenkin:1983xq,Vilenkin:1984wp,Hawking:1983hj}
when issues of initial conditions was being explored. 
Such ventures lead to 
\textit{tunnelling} proposal \cite{Vilenkin:1982de,Vilenkin:1983xq,Vilenkin:1984wp}
and \textit{no-boundary} proposal \cite{Hawking:1981gb,Hartle:1983ai,Hawking:1983hj}.
Euclidean path-integral of gravity (which is 
unbounded from below \cite{Gibbons:1977zz} due to famous conformal factor problem 
\cite{Gibbons:1978ac}) needs not only a sensible initial condition choice
but also a choice of contour of integration \cite{Halliwell:1988ik,Halliwell:1989dy,Halliwell:1990qr}.
Picard-Lefschetz theory allows one to pick the contour uniquely 
directly in Lorentzian spacetime and allows one to study
scenarios involving various initial conditions in a systematic manner
\cite{Feldbrugge:2017kzv,Feldbrugge:2017fcc,Feldbrugge:2017mbc}. 

In this paper we make use of Picard-Lefschetz theory to analyse the 
path-integral of novel-Gauss-Bonent gravity in the mini-superspace 
approximation. We ask a straightforward 
question: what is the transition probability from one state to another, 
where the states are specified by the boundary conditions
and correspond to a geometry. We seek to answer this by building on the footsteps 
of the formalism developed in \cite{Narain:2019qcj}. Due to 
complicated form of the mini-superspace action our efforts 
are limited to address the problem perturbatively in 
nGB coupling. We do the computation of transition 
amplitude to first order in nGB coupling.  

The paper has following outline: section \ref{minisup} deals with 
constructing a mini-superspace action for novel Gauss-Bonnet gravity.
Section \ref{EQM} solves the system of equations to first order in nGB coupling.
In section \ref{transAmp} we compute transition amplitude from one 
$3$-geometry to another perturbatively to first order in nGB coupling. 
Section \ref{Ninnt} deals with Picard-Lefschetz (PL) methods where beside
reviewing the PL-technology, we use it to do the integration over lapse. 
In section \ref{cuni} we study the case of classical boundary conditions 
and apply the results obtained in previous section to compute the 
transition amplitude in the case of classical Universe. We conclude 
by summarizing our findings with a discussion in section \ref{conc}.

%%%%%%%%%%%%%%%%%%%%%%%%%%%%%%%%%%%%%%%%%%%%%%%
\section{Mini-superspace action}
\label{minisup}
%%%%%%%%%%%%%%%%%%%%%%%%%%%%%%%%%%%%%%%%%%%%%%%

To compute the mini-superspace action here we first consider a 
generalization of FLRW metric in arbitrary spacetime 
dimension whose dimensionality is $D$. In polar
co-ordinates $\{t_p, r, \ta, \cdots \}$ the FLRW metric 
can be expressed as
\beq
\label{eq:frwmet}
{\rm d}s^2 = - N_p^2(t_p) {\rm d} t_p^2 
+ a^2(t_p) \left[
\frac{{\rm d}r^2}{1-kr^2} + r^2 {\rm d} \OM_{D-2}^2
\right] \, ,
\eeq
where $N_p(t_p)$ is lapse function, $a(t_p)$ is scale-factor, 
$k=(0, \pm 1)$ is the curvature, and ${\rm d}\OM_{D-2}$ is the 
metric corresponding to unit sphere in $D-2$ spatial dimensions. 
The FLRW metric is conformally related to flat metric and hence
its Weyl-tensor $C_{\mu\nu\rho\sg} =0$. For Riemann tensor 
the nonzero entries are \cite{Deruelle:1989fj,Tangherlini:1963bw,Tangherlini:1986bw} 
\bea
\label{eq:riemann}
R_{0i0j} &=& - \left(\frac{a^{\prime\prime}}{a} - \frac{a^\prime N_p^\prime}{a N_p} \right) g_{ij} \, , 
\notag \\
R_{ijkl} &=& \left(\frac{k}{a^2} + \frac{a^{\prime2}}{N_p^2 a^2} \right)
\left(g_{ik} g_{jl} - g_{il} g_{jk} \right) \, ,
\eea
where $g_{ij}$ is the spatial part of the FLRW metric
and $({}^\prime)$ denotes derivative with respect to $t_p$.
For the Ricci-tensor the non-zero components are 
\bea
\label{eq:Ricci-ten}
R_{00} &=& - (D-1) \left(\frac{a^{\prime\prime}}{a} - \frac{a^\prime N_p^\prime}{a N_p} \right)
\, , 
\notag \\
R_{ij} &=& \left[
\frac{(D-2) (k N_p^2 + a^{\prime2})}{N_p^2 a^2}
+ \frac{a^{\prime\prime} N_p - a^\prime N_p^\prime}{a N_p^3} 
\right] g_{ij} \, ,
\eea
while the Ricci-scalar for FLRW is given by
\beq
\label{eq:Ricci0}
R = 2(D-1) \left[\frac{a^{\prime\prime} N_p - a^\prime N_p^\prime}{a N_p^3} 
+ \frac{(D-2)(k N_p^2 + a^{\prime2})}{2N_p^2 a^2} \right]
\, .
\eeq
Weyl-flatness offers simplicity and allows one to express
Riemann tensor in terms of Ricci-tensor and Ricci scalar.
\bea
\label{eq:Riem_exp}
R_{\mu\nu\rho\sg} = \frac{R_{\mu\rho} g_{\nu\sg} - R_{\mu\sg}g_{\nu\rho}
+ R_{\nu\sg} g_{\mu\rho} - R_{\nu\rho} g_{\mu\sg}}{D-2}
- \frac{R (g_{\mu\rho} g_{\nu\sg} - g_{\mu\sg} g_{\nu\rho})}{(D-1)(D-2)} \,.
\eea
This identity is valid for all conformally flat metrics and allows one to express
\beq
\label{eq:Reim2_exp}
R_{\mu\nu\rho\sg} R^{\mu\nu\rho\sg}
= \frac{4}{D-2} R_\mn R^\mn - \frac{2 R^2}{(D-1)(D-2)} \, .
\eeq
By making use of this identity for conformally flat metrics in the 
Gauss-Bonnet action one can obtain a simplified action of the theory.
In such cases we have
\bea
\label{eq:actGB}
\int {\rm d}^Dx \sqrt{-g} && \left(
R_{\mu\nu\rho\sg} R^{\mu\nu\rho\sg}  - 4 R_\mn R^\mn + R^2
\right)
\notag \\
&&
= \frac{D-3}{D-2} \int {\rm d}^Dx \sqrt{-g} \left(
- R_\mn R^\mn + \frac{D R^2}{D-1}
\right) \, .
\eea
On plugging the FLRW metric of eq. (\ref{eq:frwmet}) in the action 
in eq. (\ref{eq:act}) one can get an action for $a(t_p)$ and 
$N_p(t_p)$. This action is given by,
\bea
\label{eq:midSact}
&&
S = \frac{V_{D-1}}{16 \pi G} \int {\rm d}t_p
\biggl[
a^{D-3} \biggl\{
(D-1)(D-2) k N^3 - 2 \Lam a^2 N_p^3 - 2 (D-1) a a^\prime N_p^\prime
\notag \\
&&
+ (D-1)(D-2) a^{\prime2} N_p + \underline{2 (D-1) N_p a a^{\prime\prime}}
\biggr\}
+ \frac{(D-1)(D-2)(D-3) \al}{D-4} \biggl\{
\frac{a^{D-5}(D-4)}{N_p^3} 
\notag\\
&&
\times (kN_p^2 + a^{\prime2})^2 
+ \underline{\frac{4 a^{D-4}(kN_p^2 + a^{\prime2})}{N_p^2} 
\frac{{\rm d}}{{\rm d}t_p} \left(\frac{a^\prime}{N_p}\right)}
\biggr\}
\biggr] \, ,
\eea
where $V_{D-1}$ is the volume of $D-1$ dimensional space. One can 
perform an integration by parts in the underlined terms to arrive 
at action where $D\to4$ limit can be smoothly taken. Under 
an integration by parts the $(D-4)$ factors are seen to cancel off. 
This resulting action in $D=4$ is given by,
\bea
\label{eq:act_frw_int}
S = \frac{V_3}{8 \pi G} \int {\rm d}t_p \biggl[
(3 k -  \Lam a) N_p a - \frac{3a a^\prime}{N_p} 
+ \frac{3\al}{a} \biggl\{
\frac{(kN_p^2 + a^{\prime2})^2}{N_p^3} + \frac{4k a^{\prime2}}{N_p}
+ \frac{4 a^{\prime4}}{N_p}
\biggr\}
\biggr] \, .
\eea
The Gauss-Bonnet term gives a non-trivial contribution in 
$D=4$ which is possible as its coefficient has been defined 
with a $(D-4)$ factor in denominator, which cancels off any
$(D-4)$ in numerator. With this action one can do further analysis. 
This action can be recast in to a more appealing form by a 
rescaling of lapse and scale factor. 
\beq
\label{eq:rescale}
N_p(t_p) {\rm d} t_p = \frac{N(t)}{a(t)} {\rm d} t \, ,
\hspace{5mm}
q(t) = a^2(t) \, .
\eeq
This set of transformation changes our original metric in eq. (\ref{eq:frwmet})
into following
\beq
\label{eq:frwmet_changed}
{\rm d}s^2 = - \frac{N^2}{q(t)} {\rm d} t^2 
+ q(t) \left[
\frac{{\rm d}r^2}{1-kr^2} + r^2 {\rm d} \OM_{D-2}^2
\right] \, ,
\eeq
and our action in $D=4$ given in eq. (\ref{eq:act_frw_int}) changes to following simple form.
\bea
\label{eq:Sact_frw_simp}
S = \frac{V_3}{16 \pi G} \int {\rm d}t \biggl[
(6 k - 2\Lam q) N - \frac{3 \dot{q}^2}{2N}
+ \frac{3\al}{8 N^3 q} (4k N^2 + \dot{q}^2)(4k N^2 + 5\dot{q}^2)
\biggl] \, ,
\eea
where $(\dot{})$ here represent derivative with respect to time $t$. It should be 
noticed that the action doesn't contains any derivative of $N$, which happens 
as we have performed integration by parts previously. This is an interesting 
higher-derivative action which only depends on $q$, $\dot{q}$ and $N$.

%%%%%%%%%%%%%%%%%%%%%%%%%%%%%%%%%%%%%%%%%%%%%%%
\section{Equation of motion}
\label{EQM}
%%%%%%%%%%%%%%%%%%%%%%%%%%%%%%%%%%%%%%%%%%%%%%%

The action in eq. (\ref{eq:Sact_frw_simp}) lacks any derivative term for $N$ indicating that 
variation of action with respect to $N$ will result in a 
constraint equation. Varying action with respect to $q(t)$ 
however leads to a dynamical equation for the evolution 
of $q(t)$. We choose the ADM gauge $\dot{N}=0$, which implies 
that $N(t) = N_c$ (constant). The equation of motion for $q(t)$ then 
is given by
\beq
\label{eq:eqmk0_1}
- 2 N_c \Lam + \frac{3 \ddot{q}}{N_c}
+ \frac{3\al}{8 N_c^3} \biggl[
\frac{15\dot{q}^4}{q^2} - \frac{60 \dot{q}^2 \ddot{q}}{q}
+ 24 k N_c^2 \left(\frac{\dot{q}^2}{q^2} - \frac{2 \ddot{q}}{q} \right)
- \frac{16 k^2 N_c^4}{q^2}
\biggr] = 0 \, .
\eeq
This equation contains higher-derivative contribution which is proportional to $\al$. 
It is a second order non-linear ODE. 
The higher-derivative contribution is novel here which doesn't arise if the 
Gauss-Bonnet coupling wasn't rescaled by factor of 
$(D-4)$ \cite{Glavan:2019inb}. Compared to the equation presented in 
\cite{Narain:2020qhh}, this has additional terms coming from
non-zero $k$ (non-flat Universe).
In principle one has to solve for $q(t)$ from the above equation 
for the boundary conditions 
\beq
\label{eq:bound_q}
q(t=0) = b_0 \, , 
\hspace{5mm}
q(t=1) = b_1 \, .
\eeq
One can then plug the $q(t)$-solution back into the action in eq. (\ref{eq:Sact_frw_simp}),
where we are in constant-$N$ gauge. On integrating this with respect to 
time, we arrive at the action for the constant lapse $N_c$. 
One then look for saddle points solution for 
$N_c$ which are obtained by varying this action with respect to $N_c$. 
This will be the full saddle point solution of theory. 

In practice this is not always possible. In the present case 
the evolution equation for $q(t)$ is quite complicated: higher-derivates
and non-linear. We therefore approach to solve the system perturbatively. 
We start by expanding $q(t)$ in powers of $\al$.
\beq
\label{eq:qt_exp}
q(t) = q_0(t) + \al q_1(t) + \cdots \, ,
\eeq
where $q_0$ is zeroth-order solution while $q_1$ is the first order solution. 
%
%%%%%%%%%%%%%%%%%%%%%%%%%%%%%%%%%%%%%%%%%%%%%%%
\subsection{zeroth-order}
\label{0order}
%%%%%%%%%%%%%%%%%%%%%%%%%%%%%%%%%%%%%%%%%%%%%%%

At the lowest ($\al^0$) order we have 
\beq
\label{eq:q0_evol}
\ddot{q} = \frac{2 N_c^2 \Lam}{3} \, . 
\eeq
This linear second order ODE can be solved analytically. Its solution 
obeying the boundary condition stated in eq. (\ref{eq:bound_q})
is given by
\beq
\label{eq:q0_sol}
q_0(t) = \frac{\Lam N_c^2}{3} (t^2 -t) + b_0(1-t) + b_1 t \, .
\eeq
We plug this back into the action in eq. (\ref{eq:Sact_frw_simp}) and integrate with respect to 
$t$. One gets zeroth-order action for $N_c$. This is given by
\beq
\label{eq:S0_act_N}
S_0 = \frac{V_3}{16\pi G} 
\left[-\frac{3(b_0-b_1)^2}{2N_c} + 6 k N_c
- (b_0+b_1)N_c \Lam + \frac{N_c^3 \Lam^2}{18} \right]\, .
\eeq
From the zeroth order action for $N_c$ one can compute the zeroth order 
saddle points by varying action with with respect to $N_c$. 
Then we see that $\pt S_0/\pt N_c =0$ whose solution gives $N_0$. 
\beq
\label{eq:Nc0_sad}
\frac{3(b_0-b_1)^2}{2N_0^2} 
- (b_0+b_1)\Lam + \frac{N_0^2 \Lam^2}{6} + 6 k = 0 \, .
\eeq
This is quadratic in $N_0^2$ and consist of four solutions
which are given by
\beq
\label{eq:Nc0_soln}
\left(N_0\right)_{\pm, \pm}
= \pm \sqrt{\frac{3}{\Lam}}\left(\sqrt{b_1-\frac{3k}{\Lam}} \pm \sqrt{b_0-\frac{3k}{\Lam}} \right) \, .
\eeq
At the zeroth order we don't receive any correction from the Gauss-Bonnet term 
and they agree with the known saddles in the context of Lorentzian quantum cosmology
\cite{Feldbrugge:2017kzv,Feldbrugge:2017fcc}. Corresponding to each $\left(N_0\right)_{\pm\pm}$ 
we have corresponding $\left(q_0\right)_{\pm\pm}$. Each of them leads to 
a different FLRW metric. Corresponding to each of them we have an on-Shell 
action, which is given by
\beq
\label{eq:S0act_onshell}
S_0^{\rm on-shell} = \mp \frac{V_3}{4 \pi G} \sqrt{\frac{\Lam}{3}}
\biggl[
\left(b_1 -\frac{3k}{\Lam} \right)^{3/2}
\pm 
\left(b_0 -\frac{3k}{\Lam} \right)^{3/2}
\biggr] \, .
\eeq

%%%%%%%%%%%%%%%%%%%%%%%%%%%%%%%%%%%%%%%%%%%%%%%
\subsection{First order}
\label{first order}
%%%%%%%%%%%%%%%%%%%%%%%%%%%%%%%%%%%%%%%%%%%%%%%

At first order in $\al$ the equations becomes more involved as the 
novel-Gauss Bonnet gravity starts to contribute. The evolution of 
$q(t)$ at first order is dictated by following equation 
\beq
\label{eq:q1_evo_eq}
\ddot{q}_1 = - \frac{1}{8 N_c^2} \bigg[
\frac{15\dot{q_0}^4}{q_0^2} - \frac{60 \dot{q_0}^2 \ddot{q_0}}{q_0}
+ 24 k N_c^2 \left(\frac{\dot{q_0}^2}{q_0^2} - \frac{2 \ddot{q_0}}{q_0} \right)
- \frac{16 k^2 N_c^4}{q_0^2}
\biggr] \, , 
\eeq
where $q_0$ is the zeroth order solution to $q(t)$ obtained before.
This need to be solved along with the boundary conditions for $q_1$. 
The boundary conditions for $q_1(t)$ can be obtained from eq. (\ref{eq:bound_q}) and 
those of $q_0$. This implies that 
\beq
\label{eq:boundQ1}
q_1(t=0) = q_1(t=1) = 0\, .
\eeq
The ODE for $q_1(t)$ can now be solved with these boundary conditions, and 
its solution is given by
\bea
\label{eq:q1sol}
&&
q_1(t) = \frac{5 N_c^2 \Lam^2 t(t-1)}{3} 
- \frac{1}{4 N_c^2} 
\biggl[
5 U + \frac{72 N_c^2 k}{U} - \frac{432 N_c^4 k^2}{U^3}
\biggr]
\biggl[
\left(b_0-b_1 + \frac{N_c^2\Lam}{3} \right)(t-1) 
\notag\\
&&
\times 
\tan^{-1} \left(\frac{3(b_0-b_1) + N_c^2\Lam}{U}\right)
+ \left(b_0 - b_1 - \frac{N_c^2\Lam}{3} \right)
t \tan^{-1} \left(
\frac{3(b_1-b_0) + N_c^2\Lam}{U}
\right)
\notag\\
&&
+ \biggl(b_1 - b_0 
+ \frac{N_c^2\Lam(2t-1)}{3} \biggr)
\tan^{-1} \left(\frac{3(b_1-b_0) + N_c^2 \Lam (1-2t)}{U} \right)
\biggr] \, ,
\eea
where 
\beq
\label{eq:Uform}
U = \sqrt{6(b_0+b_1)^2 N_c^2 \Lam - 9(b_0-b_1)^2 - N_c^4 \Lam^2} \, .
\eeq
Having obtained the first order correction to $q(t)$, we can plug back the 
corrected solution $q(t) = q_0 + \al q_1$ in action in eq. (\ref{eq:Sact_frw_simp})
and perform the $t$-integration. 
This results in a first order corrected action for $N_c$. 
\bea
\label{eq:S1act_N}
&&
S_1 = S_0 + \frac{V_3 \al}{16\pi G} 
\biggl[
\frac{5 (b_0 - b_1)^2 \Lam}{N_c} - \frac{5(b_0+b_1) N_c \Lam^2}{3}
+ \frac{10N_c^3 \Lam^3}{27} + 12 N_c \Lam k
+ \biggl(
\frac{5U^3}{36 N_c^3} 
\notag \\
&&
- \frac{6 U k}{N_c} + \frac{36 N_c k^2}{U}
\biggr) \biggl\{\tan^{-1} \left(\frac{3(b_0-b_1) + N_c^2\Lam}{U}\right)
+ \tan^{-1} \left(\frac{3(b_1-b_0) + N_c^2\Lam}{U}\right) \biggr\}
+ \cdots 
\biggr] \, .
\eea
This first order corrected action can be varied with respect to $N_c$ 
to obtain the correction to the zeroth order saddle points. 
To obtain this we substitute 
\beq
\label{eq:Nexp}
N_1 = N_0 + \al \nu_1 + \cdots \, .
\eeq
Then $N_1$ is the first order corrected saddle and can be obtained from  
\bea
\label{eq:N1sol_eq}
\left. \frac{\pt S_0}{\pt N_c} \right|_{N_c \to N_1= (N_0 + \al \nu_1)} = 0 \, .
\eea
On solving this equation for $\nu_1$ we get 
\bea
\label{eq:N1_sol}
&&
\nu_1 = -\frac{5N_0 \Lam}{2} 
+ \frac{k N_0^3 \Lam}{3(b_0-b_1)^2 + 6 k N_0^2 - (b_0 +b_1) N_0^2 \Lam}
\notag \\
&&
+ 4 \sqrt{k} \biggl[
\tan^{-1} \left(
\frac{3(b_0-b_1) + N_0^2\Lam}{6\sqrt{k} N_0}
\right)
+\tan^{-1} \left(
\frac{3(b_1-b_0) + N_0^2\Lam}{6\sqrt{k} N_0}
\right)
\biggr]
\, ,
\eea
where $N_0$ is given by eq. (\ref{eq:Nc0_sad}). This when combined 
with the zeroth order solution gives us the first order corrected saddles $N_1$. 
\bea
\label{eq:N1_final}
&&
N_1= N_0-\frac{5\al N_0 \Lam}{2} 
+ \frac{k \al N_0^3 \Lam}{3(b_0-b_1)^2 + 6 k N_0^2 - (b_0 +b_1) N_0^2 \Lam}
\notag \\
&&
+ 4 \al \sqrt{k} \biggl[
\tan^{-1} \left(
\frac{3(b_0-b_1) + N_0^2\Lam}{6\sqrt{k} N_0}
\right)
+\tan^{-1} \left(
\frac{3(b_1-b_0) + N_0^2\Lam}{6\sqrt{k} N_0}
\right)
\biggr]\, .
\eea
From this we can compute the first order corrected on-shell action. This is given by
\bea
\label{eq:S1_onshell}
&&
S_1^{\rm on-shell} = S_0^{\rm on-shell}
+ \frac{V_3\Lam \al}{144 G \pi} \biggl[
(b_0 - b_1)^2 - \frac{4 k N_0^2}{5} + \frac{(b_0 + b_1) N_0^2 \Lam }{3}
\biggr]
\, ,
\eea
where $S_0^{\rm on-shell}$ is the zeroth order on-shell action 
given in eq. (\ref{eq:S0act_onshell}).

%%%%%%%%%%%%%%%%%%%%%%%%%%%%%%%%%%%%%%%%%%%%%%%
\section{Transition amplitude}
\label{transAmp}
%%%%%%%%%%%%%%%%%%%%%%%%%%%%%%%%%%%%%%%%%%%%%%%

Once we have action of theory then the real important question to ask is the 
role the theory plays in quantum regimes. Such issues can only be addressed when 
one has full action of theory. In our present case it is worthy to ask the 
transition amplitude from one $3$-geometry to another. We aim to study this 
directly in Lorentzian signature by making use of WKB and Picard-Lefschetz theory
\cite{Halliwell:1988ik,Feldbrugge:2017kzv}.

The relevant quantity that we wish to compute is the transition 
probability from one $3$-geometry to another, which is a generalization 
of probability computation in usual quantum mechanics (or field theory)
to the case of gravity. In mini-superspace approximation this means
\beq
\label{eq:Gamp}
G[b_0,b_1]
= \int_{\cal C} {\cal D} N(t)  \int_{b_0}^{b_1} {\cal D} q(t) \,\, 
\exp \left(\frac{i}{\hbar} S\right) \, , 
\eeq
where $q(t)$ satisfies the boundary condition given in eq. (\ref{eq:bound_q}).
$S$ is given in eq. (\ref{eq:Sact_frw_simp}), `${\cal C}$' is the 
contour of integration for $N$ which is chosen 
using Picard-Lefschetz theory. The computation of the path-integral 
is a complicated task even in the mini-superspace approximation. The
usual complication of defining measure, convergence,
un-controllable oscillations still exist.
Often in quantum mechanical path-integral the measure is defined 
by discretising and convergence is obtained via Wick rotation.
Using Picard-Lefschetz one can generalize Feynman $+i\ep$-prescription
in a unique way thereby leading to an absolutely convergent path-integral 
along the paths of steepest descent. We will study 
this system in WKB approximation. We have already worked 
out perturbative solution to equation of motion following 
from action in eq. (\ref{eq:Sact_frw_simp}). This will be required 
in the WKB approximation, which is also gaussian approximation.

In the WKB approximation we consider fluctuation around the 
solution to equation of motion keeping the end points fixed.  
\beq
\label{eq:qdecomp}
q(t) = q_b(t) + Q(t) \, ,
\eeq
where $q_b(t)$ satisfies the equation of motion while 
$Q(t)$ is the fluctuation around the background $q_b$. We plug this in 
action given in eq. (\ref{eq:Sact_frw_simp}) and expand to second order 
in $Q(t)$. In the expansion the first order terms identically vanish 
as $q_b(t)$ satisfies equation of motion. The second order terms 
in the gauge $\dot{N}=0$ are given by,
\bea
\label{eq:S2GB_Q}
S^{(2)} &=& \frac{V_3}{2} \int_0^1 {\rm d} t
\biggl[
\underbrace{-\frac{3 \dot{Q}^2}{2N_c}}_{\rm EH}+ 
\frac{45 \al}{4N_c^3}\biggl\{
\frac{\dot{q}_b^2}{q_b} + \frac{4 k N_c^2}{5 q_b} \biggr\} \dot{Q}^2 
\notag \\
&&
+ \frac{45 \al}{8N_c^3}\biggl\{
\frac{2 \dot{q}_b^2 \ddot{q}_b}{q_b^2} - \frac{\dot{q}_b^4}{q_b^3}
+ \frac{8k N_c^2}{5} \left(
\frac{\ddot{q}_b}{q_b^2} - \frac{\dot{q}_b^2}{q_b^3} \right)
+ \frac{16 k^2 N_c^4}{15 q_b^3} 
\biggr\}Q^2 \biggl] \, ,
\eea
where we have set $(8\pi G) =1$ (which we will continue to follow from now onward)
and we have 
\beq
\label{eq:Qbound}
Q(0) = Q(1) = 0 \, .
\eeq
We have already performed integration by parts to obtain the second 
variation on eq. (\ref{eq:S2GB_Q}) which allowed us to combine certain terms.
The $q_b(t)$ entering here in the second variation is the full solution 
of the equation of motion, however in this paper we will focus on 
dealing with system to first order in $\al$. This implies that if we plug
$q_b(t) = q_0 (t) + \al q_1(t)$ in the eq. (\ref{eq:S2GB_Q}) 
(where $q_0(t)$ and $q_1(t)$ are given in eq. (\ref{eq:q0_sol})
and (\ref{eq:q1sol}) respectively) and expand to 
first order in $\al$, then we have
\bea
\label{eq:firstOD_expQ}
&&
S^{(2)} = \frac{V_3}{2} \int_0^1 {\rm d} t
\biggl[
\biggl\{-\frac{3}{2N_c}+ 
\frac{45 \al}{4N_c^3}\biggl(
\frac{\dot{q}_0^2}{q_0} + \frac{4 k N_c^2}{5 q_0} \biggr) \biggr\} \dot{Q}^2 
\notag \\
&&
+ \frac{45 \al}{8N_c^3}\biggl\{
\frac{2 \dot{q}_0^2 \ddot{q}_0}{q_0^2} - \frac{\dot{q}_0^4}{q_0^3}
+ \frac{8k N_c^2}{5} \left(
\frac{\ddot{q}_0}{q_0^2} - \frac{\dot{q}_0^2}{q_0^3} \right)
+ \frac{16 k^2 N_c^4}{15 q_0^3} 
\biggr\}Q^2 \biggl] \, ,
\eea
where the terms proportional to $\al$ constitute 
$S^{(2)}_{\rm nGB}$. 
After the decomposition we get the following form of the transition 
amplitude.
\beq
\label{eq:pathDecomp}
G[b_0,b_1]
= \int_{0^+}^{\infty} {\rm d} N_c \exp\left( \frac{iS_1}{\hbar}\right)
\int_{Q[0]=0}^{Q[1]=0} {\cal D} Q(t) \exp\left(\frac{i S^{(2)}}{\hbar}\right) \, ,
\eeq
where $S_1$ is given in eq. (\ref{eq:S1act_N}). Our task first then is to compute the 
path-integral over $Q(t)$. In the case when the second variation contains only terms 
coming from Einstein-Hilbert gravity ($\al\to0$), then the $Q$-path-integral is easy 
to perform exactly. For nonzero $\al$ this is complicated as the 
coefficient of $\dot{Q}^2$ and $Q^2$ are $t$-dependent functions.
However, to first order in $\al$ one can perform path-integral perturbatively, 
which is what we will do. We will closely follow the strategy outlined in 
\cite{Narain:2019qcj}.

%%%%%%%%%%%%%%%%%%%%%%%%%%%%%%%%%%%%%%%%%%%%%%%
\subsection{$Q$-integration}
\label{Qint}
%%%%%%%%%%%%%%%%%%%%%%%%%%%%%%%%%%%%%%%%%%%%%%%

We first note that in the second variation the coefficient of $\dot{Q}^2$ 
and $Q^2$ has time dependence, which arise as the terms proportional 
to $\al$ depend of $q_0(t)$ and its derivatives. Although $q_0(t)$ is a simple 
quadratic polynomial in $t$, it still makes it tricky to evaluate the 
path-integral exactly. We note that as fluctuation $Q(t)$ vanishes at 
the two boundary points, it implies that it has following decomposition
\beq
\label{eq:Qdecomp}
Q(t) = \sum_{\lvert k \rvert \geq 1}^{\infty} c_k \exp \left(2 \pi k t \right) \, ,
\hspace{5mm}
{\rm with}
\hspace{5mm}
c_{-k} = c^*_k \, .
\eeq
The path-integral measure accordingly becomes the 
following
\beq
\label{eq:meaDecomp}
{\cal D} Q(t) = {\cal N} \int_{-\infty}^{\infty} \prod_{\lvert k \rvert \geq 1}^{\infty} {\rm d} c_k \, ,
\eeq
where the normalization ${\cal N}$ needs to be fixed carefully. 
Usually the normalization is fixed in such a way so that it 
absorbs the infinities coming from the infinite-product or infinite 
summation. In the case of Einstein-Hilbert gravity (setting $\al\to0$ above)
we have the following $Q$-path-integral. 
\beq
\label{eq:pathQEH}
\int_{Q[0]=0}^{Q[1]=0} {\cal D} Q(t) \exp\left(
-\frac{3 i V_3}{4N_c \hbar} \int_{0}^1 {\rm d} t \dot{Q}^2
\right) 
= \left(
\frac{3 i V_3}{4 \pi N_c \hbar}
\right)^{1/2}\, .
\eeq
The expression on LHS is similar to path-integral of free particle. 
It can be evaluated exactly and it has a finite value on RHS. If we 
insert the decomposition of $Q(t)$ from 
eq. (\ref{eq:Qdecomp}) and write the measure as in eq. (\ref{eq:meaDecomp}),
then by performing the path-integral one encounters infinities. 
\beq
\label{eq:meaEH}
{\cal N}_{\rm EH} \int_{-\infty}^{\infty} 
\prod_{\lvert k \rvert \geq 1}^{\infty} {\rm d} c_k
\exp \left(
\frac{3 i V_3}{4 N_c \hbar} \sum_k (2\pi k)^2 \lvert c_k \rvert^2
\right)
= \left(
\frac{3 i V_3}{4 \pi N_c \hbar}
\right)^{1/2} \, .
\eeq
The infinity arising from the infinite-product on the LHS 
can be absorbed by suitably defining ${\cal N}_{\rm EH}$. This will give  
\beq
\label{eq:Neh_exp}
{\cal N}_{\rm EH} \prod_{k=1}^{\infty} \frac{8 \pi N_c \hbar}{3 (2 \pi k)^2 V_3} 
= \left(
\frac{3 i V_3}{4 \pi N_c \hbar}
\right)^{1/2} \, .
\eeq
For the case of novel-GB gravity, the normalisation needs to be 
fixed accordingly. At this point it is best we also write $c_n = a_n + i b_n$
and $c_{-n} = c_n^* = a_n - i b_n$, where $a_n$ and $b_n$ are real numbers.
Such a change of variables will lead to a Jacobian factor. 
The gravity action here consists of two parts: $S^{(2)}
= S^{(2)}_{\rm EH} + S^{(2)}_{\rm nGB}$. 
To first order in $\al$ we then have 
\bea
\label{eq:nGBpath_exp}
&&
\int_{Q[0]=0}^{Q[1]=0} {\cal D} Q(t) \exp\left(\frac{i S^{(2)}}{\hbar}\right)
= {\cal N}_{\rm EH} \left(1 + \al {\cal N}_1 + \cdots \right)
\notag \\
&&
\times
\int_{-\infty}^{\infty} \prod_{k=1}^{\infty} {\rm d} a_k {\rm d} b_k \left(\frac{2}{i}\right)
\left(1 + \frac{i}{\hbar} S^{(2)}_{\rm nGB} + \cdots \right)
\exp\left(\frac{i S^{(2)}_{\rm EH}}{\hbar}\right) \, ,
\notag \\
&&
= {\cal N}_{\rm EH} \int_{-\infty}^{\infty} \prod_{k=1}^{\infty} {\rm d} a_k {\rm d} b_k \left(\frac{2}{i}\right)
\exp\left(\frac{i S^{(2)}_{\rm EH}}{\hbar}\right)
\notag \\
&&
+ {\cal N}_{\rm EH} \int_{-\infty}^{\infty} \prod_{k=1}^{\infty} {\rm d} a_k {\rm d} b_k \left(\frac{2}{i}\right)
\biggl[
\al {\cal N}_1+ \frac{i}{\hbar} S^{(2)}_{\rm nGB} 
\biggr] \exp\left(\frac{i S^{(2)}_{\rm EH}}{\hbar}\right)
+ {\cal O}(\al^2) \, ,
\eea
where ${\cal N}_{\rm EH}$ is given in eq. (\ref{eq:Neh_exp}), 
${\cal N}_1$ is the infinite constant which will be adjusted to 
absorb the infinity coming from novel Gauss-Bonnet gravity part, 
while the factor $2/i$ arises due to Jacobian transformation. 
The EH action is quadratic in $a_n$ and $b_n$, which is 
easy to see once we plug the decomposition 
for $Q(t)$ and integrate with respect to time. 
$S^{(2)}_{\rm nGB}$ on the other hand contains mixed terms.
For example terms like $a_m a_n$, $a_m b_n$, and $b_m b_n$ 
(where $m$ and $n$ need not be the same) occur.
Such kind of terms don't disappear even after the $t$-integration. 
The $S^{(2)}_{\rm nGB}$ is given by,
\bea
\label{eq:nGBS2_cns}
&&
S^{(2)}_{\rm nGB} = \frac{45 \al V_3}{8 N_c^3}  \sum_{\lvert k, k^\prime \rvert \geq 1}^{\infty}
\int_0^1 {\rm d}t 
\biggl[
\biggl(
\frac{\dot{q}_0^2}{q_0} + \frac{4 k N_c^2}{5 q_0} \biggr) (4\pi^2 k k^\prime)
\notag \\
&&
+\biggl\{
\frac{\dot{q}_0^2 \ddot{q}_0}{q_0^2} - \frac{\dot{q}_0^4}{2q_0^3}
+ \frac{4k N_c^2}{5} \left(
\frac{\ddot{q}_0}{q_0^2} - \frac{\dot{q}_0^2}{q_0^3} \right)
+ \frac{8 k^2 N_c^4}{15 q_0^3} 
\biggr\}
\biggr] c_k c_{k^\prime} e^{2\pi i(k+k^\prime)t} \, ,
\eea
where $q_0(t)$ is quadratic in $t$ and is given in eq. (\ref{eq:q0_sol}). 
Here we need to perform $t$-integration. 
On plugging decomposition of $c_k$'s in terms of $a_k$'s and 
$b_k$'s, it is possible to write the above expression as a summation 
over only positive integer values of $k$ and $k^\prime$. The resulting 
expression will also contain mixed terms which are non-diagonal.
We introduce a shorthand 
\bea
\label{eq:Mkkshort}
&&
M(k,k^\prime) =  \frac{45 \al }{4 N_c^3}
\int_0^1 {\rm d}t 
\biggl[
\biggl(
\frac{\dot{q}_0^2}{q_0} + \frac{4 k N_c^2}{5 q_0} \biggr) (4\pi^2 k k^\prime)
\notag \\
&&
+\biggl\{
\frac{\dot{q}_0^2 \ddot{q}_0}{q_0^2} - \frac{\dot{q}_0^4}{2q_0^3}
+ \frac{4k N_c^2}{5} \left(
\frac{\ddot{q}_0}{q_0^2} - \frac{\dot{q}_0^2}{q_0^3} \right)
+ \frac{8 k^2 N_c^4}{15 q_0^3} 
\biggr\}
\biggr] e^{2\pi i(k+k^\prime)t} 
\notag\\
&&
= \int_0^1 {\rm d}t 
\left[
4\pi^2 k k^\prime A_1(t) +  A_2(t) 
\right] e^{2\pi i(k+k^\prime)t} \, ,
\eea
where we have 
\begin{align}
\label{eq:A1}
&
A_1(t)= \frac{45\al}{4N_c^3} \biggl(
\frac{\dot{q}_0^2}{q_0} + \frac{4 k N_c^2}{5 q_0} \biggr) \, , \\
\label{eq:A2}
&
A_2(t) = \frac{45\al}{4N_c^3} \biggl\{
\frac{\dot{q}_0^2 \ddot{q}_0}{q_0^2} - \frac{\dot{q}_0^4}{2q_0^3}
+ \frac{4k N_c^2}{5} \left(
\frac{\ddot{q}_0}{q_0^2} - \frac{\dot{q}_0^2}{q_0^3} \right)
+ \frac{8 k^2 N_c^4}{15 q_0^3} 
\biggr\}   \, .
\end{align}
This shorthand is useful as it expresses the structure of the 
$S^{(2)}_{\rm nGB}$ in a simple manner. This is given by,
\bea
\label{eq:S2nGBshort}
&&
\hspace{-10mm}
S^{(2)}_{\rm nGB} = \frac{V_3}{2}  \sum_{k, k^\prime \geq 1}^{\infty}
\biggl[
\left(M(k,k^\prime) + M(k,-k^\prime)+ M(-k,k^\prime)+ M(-k,-k^\prime) \right) a_k a_{k^\prime}
\notag\\
&&
\hspace{-10mm}
+ i\left(M(k,k^\prime) - M(k,-k^\prime)+ M(-k,k^\prime)- M(-k,-k^\prime) \right) a_k b_{k^\prime}
\notag\\
&&
\hspace{-10mm}
+ i \left(M(k,k^\prime) + M(k,-k^\prime)- M(-k,k^\prime)- M(-k,-k^\prime) \right) b_k a_{k^\prime}
\notag\\
&&
\hspace{-10mm}
+ \left(-M(k,k^\prime) + M(k,-k^\prime) - M(-k,k^\prime)+ M(-k,-k^\prime) \right) b_k b_{k^\prime}
\biggr] \, ,
\notag \\
&&
\hspace{-10mm}
= \frac{V_3}{2}  \sum_{k, k^\prime \geq 1}^{\infty}
\biggl[ 
M_{11}(k,k^\prime) a_k a_{k^\prime}
+ M_{12} (k,k^\prime) a_k b_{k^\prime}
+ M_{21} (k,k^\prime) b_k a_{k^\prime}
+ M_{22} (k,k^\prime) b_k b_{k^\prime} 
\biggr] \, .
\eea
In the path-integral given in eq. (\ref{eq:nGBpath_exp}) one has to take 
expectation value of $S^{(2)}_{\rm nGB}$. We notice the occurrence of
mixed terms in $S^{(2)}_{\rm nGB}$ given in eq. (\ref{eq:S2nGBshort}).
These non-diagonal terms don't contribute as the 
action appearing in exponent is quadratic in $a_k$ and $b_k$. 
As a result only $M_{11}(k,k^\prime)$ and $M_{22}(k,k^\prime)$
contributes. Also, among them the non-vanishing contribution 
comes only when $k=k^\prime$. 
These observations simplify our perturbative computations drastically. 
For $k=k^\prime$ the expressions for $M_{11}(k,k)$ and $M_{22}(k,k)$
is given by,
\begin{align}
\label{eq:M11}
M_{11}(k,k) &= \int_0^1 {\rm d}t
\biggl[
\left\{A_1 (2\pi k)^2 + A_2\right\} 2 \cos (4\pi k t)
+2 \left\{-A_1 (2\pi k)^2 + A_2 \right\} \biggr] \, ,
\\
M_{22}(k,k) & = \int_0^1 {\rm d}t 
\biggl[
- \left\{A_1 (2\pi k)^2 + A_2\right\} 2 \cos (4\pi k t)
+ 2 \left\{-A_1 (2\pi k)^2 + A_2 \right\} \biggr]\, .
\end{align}
Achieving great simplification we now only need to perform the integrations over 
$a_k$ and $b_k$ as dictated by the path-integral in eq. (\ref{eq:nGBpath_exp}). 
This path-integral has two parts: the leading piece is the 
Einstein-Hilbert piece which has been computed before in eq. (\ref{eq:meaEH}) 
while the second term is the correction term coming from the nGB. 
We will compute this piece now. Performing the integrations over 
$a_k$ and $b_k$, and making use of eq. (\ref{eq:Neh_exp}) we get
the following
\bea
\label{eq:akbkint}
&&
{\cal N}_{\rm EH} \int_{-\infty}^{\infty} \prod_{k=1}^{\infty} {\rm d} a_k {\rm d} b_k \left(\frac{2}{i}\right)
\biggl[
\al {\cal N}_1+ \frac{i}{\hbar} S^{(2)}_{\rm nGB} 
\biggr] \exp\left(\frac{i S^{(2)}_{\rm EH}}{\hbar}\right) 
\notag \\
&&
= \left(
\frac{3 i V_3}{4 \pi N_c \hbar}
\right)^{1/2} \biggl[
\al {\cal N}_1 - \frac{N_c}{3} \sum_{k=1}^\infty
\left\{
M_{11}(k,k) + M_{22}(k,k)
\right\} (2 \pi k)^{-2}
\biggr] \, ,
\notag \\
&&
= \left(
\frac{3 i V_3}{4 \pi N_c \hbar}
\right)^{1/2} \biggl[
\al {\cal N}_1 - \frac{4 N_c}{3} \sum_{k=1}^\infty
\int_0^1 {\rm d}t 
\left\{
A_1 + A_2 (2 \pi k)^{-2}
\right\} 
\biggr] \, ,
\notag \\
&&
= - \frac{N_c}{18}
\left(\frac{3 i V_3}{4 \pi N_c \hbar}\right)^{1/2} \int_0^1 {\rm d}t A_2 \, ,
\eea
where we have absorbed the infinite piece by defining 
the infinite constant ${\cal N}_1$ as
\beq
\label{eq:N1fixed}
{\cal N}_1 = \frac{4 N_c}{3\al} \sum_{k=1}^\infty
\int_0^1 {\rm d}t  A_1\, ,
\eeq
and $A_2(t)$ is given in eq. (\ref{eq:A2}). Putting together 
all terms we find the value of the $Q$-integration to be
\beq
\label{eq:Qint_end}
\int_{Q[0]=0}^{Q[1]=0} {\cal D} Q(t) \exp\left(\frac{i S^{(2)}}{\hbar}\right)
= \left(\frac{3 i V_3}{4 \pi N_c \hbar}\right)^{1/2} 
\biggl[
1 - \frac{N_c}{18} \int_0^1 {\rm d}t A_2 + {\cal O}(\al^2)
\biggr] \, .
\eeq
The $t$-integration here $A_2$ can be performed using {\it Mathematica}. 
It carries crucial $N_c$ dependence 
which is important in the $N_c$-integration of the path-integral
for transition amplitude.
We now have the relevant ingredients 
necessary to write an expression for the transition 
probability. We plug them in eq. (\ref{eq:pathDecomp}), which gives
\beq
\label{eq:Gab_exp}
G[b_0,b_1]
= \int_{0^+}^{\infty} {\rm d} N_c \exp\left( \frac{iS_{1}}{\hbar}\right)
\left(\frac{3 i V_3}{4 \pi N_c \hbar}\right)^{1/2} 
\biggl[
1 - \frac{N_c}{18} \int_0^1 {\rm d}t A_2
+ {\cal O}(\al^2)
\biggr] \, ,
\eeq
where the form of $A_2(t)$ is given in eq. (\ref{eq:A2})
and $S_1$ is given in eq. (\ref{eq:S1act_N}) 
After performing the $t$-integration one obtains $A_2$. 
This is given by
\bea
\label{eq:IntA2t}
&&
\hspace{-5mm}
\int_0^1 {\rm d}t A_2 = 
\frac{\al \{3(b_0+b_1) - N_c^2 \Lam\}}{48 b_0^2 b_1^2 N_c^3}
\bigl[
5(b_0 -b_1) U^2 -72 (b_0^2 + b_1^2) k N_c^2 
-10 N_c^4 \Lam^2 b_0 b_1
\bigr]
\notag \\
&&
+ \frac{9 k^2 \al N_c}{b_0^2 b_1^2 U^4}
\bigl[
(b_0^2 + b_1^2) k N_c^2 - 3 (b_0+b_1)(b_0^2 + 4 b_0 b_1 + b_1^2)
\bigr]
+ \frac{27 \al k (U^2 + 18 k N_c^2)}{b_0 b_1 N_c U^4}
\notag \\
&&
\times 
\bigl[
(b_0^2 + 6 b_0 b_1 + b_1^2) N_c^2 \Lam 
- 3 (b_0 - b_1)^2 (b_0 + b_1)
\bigr] 
- \frac{3 \al \Lam^2 N_c}{2U} \left(5 + \frac{36 N_c^2 k}{U^2}
+ \frac{648 N_c^4 k^2}{U^4} \right)
\notag \\
&&
\times
 \biggl\{\tan^{-1} \left(\frac{3(b_0-b_1) + N_c^2\Lam}{U}\right)
+ \tan^{-1} \left(\frac{3(b_1-b_0) + N_c^2\Lam}{U}\right) \biggr\}
\, ,
\eea
where $U$ is given in eq. (\ref{eq:Uform}). Our $S_1$ in eq .(\ref{eq:S1act_N})
consist of two parts: Einstein-Hilbert piece and a first order correction piece
coming from nGB.
The integration over $N_c$ has to 
be performed carefully as the integrand has singularity at $N_c=0$. In the 
complex $N_c$ plane the integrand has a branch-cut along the positive 
real axis. We will use Picard-Lefschetz theory to study this integration 
in the complex plane.

%%%%%%%%%%%%%%%%%%%%%%%%%%%%%%%%%%%%%%%%%%%%%%%
\section{Picard-Lefschetz and $N_c$-integration}
\label{Ninnt}
%%%%%%%%%%%%%%%%%%%%%%%%%%%%%%%%%%%%%%%%%%%%%%%

Our task then reduces to the computation of the $N_c$-integration,
which will be studied in complex plane. We make use of 
complex analysis and methods of Picard-Lefschetz theory 
\cite{Witten:2010cx,Witten:2010zr,Basar:2013eka,Tanizaki:2014xba}. 
In the complex plane we work out steepest descent/ascent paths 
which allow us to determined the relevant contours of integration. 
We then sum over the contribution of all such paths to find the 
transition amplitude. This powerful methodology offers a natural 
exponential damping along each thimble instead of an oscillatory integral. 

To describe the process we start with the following generic path-integral 
\beq
\label{eq:pathmock}
I = \int {\cal D}z(t) \, e^{i {\cal S}(z)/\hbar} \, ,
\eeq
where the exponent is a functional of $z(t)$. In situations when the action
${\cal S}(z)$ becomes large, then the integrand starts to oscillate 
violently. In flat spacetime field theory the usual strategy to tame such 
behavior is to Wick rotate the integration contour. This transforms the 
oscillatory integral into an exponentially damped integral. In PL-theory 
one lifts both $z$ and ${\cal S}$ in complex plane where one 
interprets ${\cal S}$ as an holomorphic functional of
$z(t)$ satisfying a functional form of Cauchy-Riemann 
conditions
\begin{align}
\label{eq:CRfunc}
\frac{\de {\cal S}}{\de \bar{z}} = 0 
\Rightarrow
\begin{cases}
\frac{\de {\rm Re} {\cal S}}{\de x}
&= \frac{ \de {\rm Im} {\cal S}}{\de y} \, , \\
\frac{\de {\rm Re} {\cal S}}{\de y}
&= - \frac{ \de {\rm Im} {\cal S}}{\de x} \, .
\end{cases}
\end{align}
%

%%%%%%%%%%%%%%%%%%%%%%%%%%%%%%%%%%%%%%%%%%%%%%%
\subsection{Flow equations}
\label{floweq}
%%%%%%%%%%%%%%%%%%%%%%%%%%%%%%%%%%%%%%%%%%%%%%%

Writing the complex exponential as
${\cal I} = i {\cal S}/\hbar = h + iH$ and 
$z(t) = x_1(t) + i x_2(t)$ then evolution downstream 
is defined as
\beq
\label{eq:downFlowDef}
\frac{{\rm d} x_i}{{\rm d} \lam}
= - g_{ij} \frac{\pt h}{\pt x_j} \, ,
\eeq
where $g_{ij}$ is a metric defined on the complex manifold, 
$\lam$ is flow parameter and $(-)$ sign refers to downward flow. 
These are the steepest descent 
contours also knowns as \textit{thimbles} and denoted by ${\cal J}_\sg$.
Steepest ascent contours are defined by plus sign in front of 
$g_{ij}$ in the above equation, and are denoted as ${\cal K}_\sg$. Here 
$\sg$ refers to the saddle point to which it is attached.
This definition automatically implies that the real part $h$ (also called 
Morse function) decreases monotonically along the steepest descent 
contour as one moves away from the critical point along the flows.
This can be seen by computing 
\beq
\label{eq:flowMonoDec_h}
\frac{{\rm d} h}{{\rm d} \lam}
= g_{ij} \frac{{\rm d} x^i}{{\rm d} \lam} \frac{\pt h}{\pt x_j} 
= - \left(\frac{{\rm d} x_i}{{\rm d}\lam}\frac{{\rm d} x^i}{{\rm d}\lam}\right)
\leq 0 \, .
\eeq
It holds generically for any Riemannian metric. However, for 
simplicity we can consider $g_{z,z}=g_{\bar{z},\bar{z}}=0$
and $g_{z,\bar{z}}= g_{\bar{z},z}=1/2$. This leads to simplified version of
flow equations 
\beq
\label{eq:simpflow}
\frac{{\rm d}z}{{\rm d} \lam} = \pm \frac{\pt \bar{\cal I}}{\pt \bar{z}} \, ,
\hspace{5mm}
\frac{{\rm d}\bar{z}}{{\rm d} \lam} = \pm \frac{\pt {\cal I}}{\pt z} \, .
\eeq
An immediate outcome of these flow equations is that the imaginary 
part of ${\rm Im} {\cal I}=H$ is constant along all the flow lines. 
\beq
\label{eq:consHflow}
\frac{{\rm d}H}{{\rm d} \lam} 
= \frac{1}{2i} \frac{{\rm d} ({\cal I} - {\cal \bar{I}})}{{\rm d} \lam} 
= \frac{1}{2i} \left(
\frac{\pt {\cal I}}{\pt z} \frac{{\rm d} z}{{\rm d} \lam} 
- \frac{\pt \bar{\cal I}}{\pt \bar{z}}\frac{{\rm d}\bar{z}}{{\rm d} \lam}
\right) = 0 \, .
\eeq
This is a wonderful feature of flow-lines which can be exploited 
to determine them quickly. In the complex $N_c$-plane, in cartesian 
co-ordinates language, the flow equations corresponding to 
steepest descent (ascent) becomes the following 
\begin{subequations}
\begin{align}
\label{eq:STdes}
& {\rm Descent} \Rightarrow 
& \frac{{\rm d} x_1}{{\rm d} \lam} = - \frac{\pt {\rm Re}{\cal I}}{\pt x_1} \, ,
\hspace{5mm}
&
\frac{{\rm d} x_2}{{\rm d} \lam} = - \frac{\pt {\rm Re}{\cal I}}{\pt x_2} \, , 
\\
\label{eq:STaes}
& {\rm Ascent} \Rightarrow 
& \frac{{\rm d} x_1}{{\rm d} \lam} = \frac{\pt {\rm Re}{\cal I}}{\pt x_1} \, ,
\hspace{5mm}
&
\frac{{\rm d} x_2}{{\rm d} \lam} = \frac{\pt {\rm Re}{\cal I}}{\pt x_2} \, ,
\end{align}
\end{subequations}
as the $\de \left({\rm Im} {\cal I}\right)= 0$ along the flow lines. These equations can be used 
to determine the trajectories of the 
steepest descent and ascent in the complex $N_c$-plane emanating
from the saddle point. Each saddle point has a steepest descent trajectory 
starting from it and a steepest ascent trajectory ending in it. 
Based on boundary condition and the values of various parameters,
the location of saddles move accordingly. Similarly the behavior of trajectories 
and their shape also changes. Usually these equations are coupled 
ODEs and can be complicated to solve analytically in complicated 
system like in present case. These flow lines can also be determined by exploiting the 
knowledge that $H$ is constant along them, however to determine the 
ascent/descent one needs to compute the gradient of first derivative 
(second derivative at the saddle points). 

%%%%%%%%%%%%%%%%%%%%%%%%%%%%%%%%%%%%%%%%%%%%%%%
\subsection{Choice of contour}
\label{choice}
%%%%%%%%%%%%%%%%%%%%%%%%%%%%%%%%%%%%%%%%%%%%%%%

Once the set of steepest descent/ascent trajectories and saddle points are known, it 
can be used to determine the contour of integration in the complex $N_c$-plane.
This is the deformed contour of integration to which the original contour is deformed. 
Along this new path of integration the $N_c$ integral becomes absolutely convergent
as discussed in great detail in \cite{Feldbrugge:2017kzv}. However, determining a 
suitable path of contour need some work. Part of the job is done once steepest 
descent ${\cal J}_\sg$ and ascent paths ${\cal K}_\sg$, and saddle points $N_\sg$
are known. 

In the complex $N_c$ plane one can study the behavior of $h$ and $H$, and 
determine the allowed region (region where integral is well-behaved) and 
forbidden region (region where integral diverges). The former is 
denoted by $J_\sg$ while later is denoted by $K_\sg$. It is seen 
that $h(J_\sg) < h(N_\sg)$, while $h(K_\sg)>h(N_\sg)$. 
Moreover, generically it is seen that $h$ goes 
to $-\infty$ along the steepest descent lines and ends in singularity, 
while steepest ascent contours end in singularity where $h\to+\infty$. 
These two lines intersect at only one point where they are both 
well-defined. Our task is to choose a contour of integration which lies 
in region $J_\sg$ and follows along the steepest descent paths \cite{Feldbrugge:2017kzv}.
The relevance of saddle is decided when the steepest ascent path emanating 
from it intersects the original path of integration. The Lefschetz thimble passing 
though this saddle point becomes the relevant ${\cal J}_\sg$, as the intersection 
point of ${\cal J}_\sg$ and ${\cal K}_\sg$ smoothly moves over 
to the intersection of ${\cal K}_\sg$ with original contour. Thus
cleanly deforming the original contour to path along Lefschetz thimbles.

Then the original integration $(0^+,\infty)$ is a summation over 
contribution from all the steepest descent contours passing through relevant saddles. 
Formally it can be expressed as
\beq
\label{eq:sadsum}
(0^+,\infty) = \sum_\sg n_\sg {\cal J}_\sg \, ,
\eeq
where $n_\sg$ takes values $\pm1, 0$ depending on the relevance of 
saddles, while ${\cal J}_\sg$ here refers to integration performed 
along the steepest descent path. Once we have deformed the 
contour from the original integration path to sum over various 
relevant thimbles we have 
\beq
\label{eq:sumOthim}
I = \int_{\cal C} {\rm d} z(t) e^{i S[z]/\hbar}
= \sum_\sg n_\sg \int_{{\cal J}_\sg} {\rm d}z(t)
e^{i S[z]/\hbar} \, .
\eeq
Usually more than one thimbles contribute leading to an occurrence of
an interference. This is the Lorentzian path integral
which is summation of contribution from various relevant thimbles. 
The integration over each thimble is absolutely convergent if
\beq
\label{eq:absconvg}
\biggl| \int_{{\cal J}_\sg} {\rm d} z(t)
e^{i S[z]/\hbar} \biggr| \leq
\int_{{\cal J}_\sg} \lvert {\rm d} z(t) \rvert
\lvert e^{i S[z]/\hbar} \rvert
= \int_{{\cal J}_\sg} \lvert {\rm d} z(t) \rvert e^h(z) < \infty \, .
\eeq
Defining the length along the curve as $l= \int \lvert {\rm d} z(t) \rvert$,
then the above integral is convergent if $e^h \sim 1/l$ as $l\to \infty$. 
Then the original integration becomes a sum of absolutely 
convergent steepest descent integrals. On doing an expansion in 
$\hbar$ we get the following leading order piece 
\beq
\label{eq:LDordI}
I = \int_{\cal C} {\rm d} z(t) e^{i S[z]/\hbar}
= \sum_\sg n_\sg e^{i H(N_\sg)} \int_{{\cal J}_\sg} {\rm d}z(t) e^{h} 
\approx \sum_\sg n_\sg e^{i S[N_\sg]/\hbar} \left[A_\sg 
+ {\cal O}(\hbar) \right] \, ,
\eeq
where $A_\sg$ is the contribution coming after doing a 
gaussian integration around the saddle point $N_\sg$.

%%%%%%%%%%%%%%%%%%%%%%%%%%%%%%%%%%%%%%%%%%%%%%%
\subsection{Flow directions}
\label{flowDir}
%%%%%%%%%%%%%%%%%%%%%%%%%%%%%%%%%%%%%%%%%%%%%%%

The flow-directions can be determined by computing second derivative 
of action with respect to $N_c$ at the saddle points.
Writing $N_c = N_\sg + \de N$ (where $N_\sg$ is any saddle point of action), the
 action has a power series expansion in $\de N$.
\beq
\label{eq:NexpSad}
S^{(0)} = S^{(0)}_s + \left. \frac{{\rm d}S^{(0)}}{{\rm d}N_c} \right|_{N_c=N_\sg} \de N_c 
+ \frac{1}{2} \left. \frac{{\rm d}^2 S^{(0)}}{{\rm d}N_c^2} \right|_{N_c=N_\sg} \left(\de N_c\right)^2 
+ \cdots
\eeq
The first order terms vanish identically as $N_\sg$ are saddle points. 
The second order terms can be obtained directly from the action 
in eq. (\ref{eq:S1act_N}) by taking double derivative with respect to $N_c$.
From this the direction of flows can be determined. One should remember that 
the imaginary part of exponential $H$ is constant along the flow lines.
This immediately leads to ${\rm Im} \left[iS - i S (N_\sg) \right]=0$.
The second variation at the saddle point can be expressed 
as a complex number ${\rm d}^2 S^{(0)}/{\rm d}N^2 = re^{i \rho}$, where 
$r$ and $\rho$ depends on boundary conditions. 
Near the saddle point the change in $H$ will go like 
\beq
\label{eq:changeH}
\D(H) \propto i 
\left(\left. \frac{{\rm d}^2 S^{(0)}}{{\rm d}N^2} \right|_{N_\sg}\right) \left(\de N_c\right)^2
\sim n^2 e^{i\left(\pi/2 + 2\ta + \rho \right)} \, ,
\eeq
where we have written $\de N_c = n e^{i \ta}$ and $\ta$ is the direction of flow lines. 
As the imaginary part of $H$ remains constant along the 
flow lines, so this implies
\beq
\label{eq:flowang}
\ta = \frac{(2k-1)\pi}{4} - \frac{\rho}{2} \, ,
\eeq
where $k \in \mathbb{Z}$. The steepest descent/ascent flow lines 
have angles $\ta_{\rm des/aes}$ respectively, where the phase for $\D H$ is such 
that it correspond to $e^{i\left(\pi/2 + 2\ta + \rho \right)} = \mp1$. This implies
\beq
\label{eq:TaDesAes}
\ta_{\rm des} = k \pi + \frac{\pi}{4} - \frac{\rho}{2} \, ,
\hspace{5mm}
\ta_{\rm aes} = k \pi - \frac{\pi}{4} - \frac{\rho}{2} \, .
\eeq
These angles can be computed numerically for the given boundary conditions
and for gravitational actions.

%%%%%%%%%%%%%%%%%%%%%%%%%%%%%%%%%%%%%%%%%%%%%%%
\subsection{Saddle-point approximation}
\label{sadPtApp}
%%%%%%%%%%%%%%%%%%%%%%%%%%%%%%%%%%%%%%%%%%%%%%%

Once we have the information about the saddles, flow-lines, their directions, 
and steepest descent/ascent paths 
(denoted by ${\cal J}_\sg/{\cal K}_\sg$ respectively), it is then easy to figure 
out the relevant saddle points.
When the steepest ascent path emanating from a saddle point
coincides with the original contour of integration (which in this case is
$(0^+, \infty)$), then it is a relevant saddle point. The original 
integration contour then becomes sum over the contribution 
coming from all the Lefschetz thimbles through relevant saddle points.
The path-integral giving transition amplitude in eq. (\ref{eq:Gab_exp}) 
then becomes following 
\bea
\label{eq:Gabapprox}
&&
G[b_0,b_1] = \sum_\sg n_\sg 
\left(\frac{3 i V_3}{4 \pi \hbar}\right)^{1/2} \int_{{\cal J}_\sg} \frac{{\rm d}N_c}{\sqrt{N_c}}
\exp\left( \frac{iS_1}{\hbar}\right)
\biggl[
1 - \frac{N_c}{18} \int_0^1 {\rm d}t A_2
+ {\cal O}(\al^2)
\biggr] \, ,
\notag \\
&&
\approx
\sum_\sg n_\sg 
\left(\frac{3 i V_3}{4 \pi \hbar}\right)^{1/2}
\exp\left( \frac{iS_1^{\rm on-shell}}{\hbar}\right)
\frac{1}{\sqrt{N_\sg}}
\biggl[
1 - \frac{N_\sg}{18} \int_0^1 {\rm d}t A_2 (N_\sg)
+ {\cal O}(\al^2) 
\biggr] 
\notag \\
&&
\times \int_{{\cal J}_\sg}
{\rm d}N_c \exp \left(
\frac{i \left(S_1\right)_{N_cN_c}}{\hbar} (N_c - N_\sg)^2
\right)\biggl(1 + {\cal O}(\sqrt{\hbar}) \biggr) \, ,
\eea
where $N_\sg$ are saddle points which to first order in $\al$ are given in 
eq. (\ref{eq:N1_final}), $S_1$ is given in eq. (\ref{eq:S1act_N}), 
$S_1^{\rm on-shell}$ is given in eq. (\ref{eq:S1_onshell}) 
and $A_2(N_\sg)$ can be computed from eq. (\ref{eq:IntA2t}). The second variation 
of action with respect to $N_c$ computed at the saddle point 
and to first order in $\al$ is given by following
\bea
\label{eq:S0NNsadpt}
&&
\left. \left(S_1\right)_{N_cN_c} \right|_{N_\sg} = 
\frac{V_3}{2} \biggl[
- \frac{(2+3\al) \left\{3(b_0 -b_1)^2 + 6 kN_0^2 - (b_0+b_1) N_0^2 \Lam \right\}}{2 N_0^3}
\notag \\
&&
+ \frac{k \al \Lam \left\{ 3(b_0 -b_1)^2 - 6 kN_0^2 + (b_0+b_1) N_0^2 \Lam\right\}}
{N_0 \left\{3(b_0 -b_1)^2 + 6 kN_0^2 - (b_0+b_1) N_0^2 \Lam\right\} }
- \frac{24 \al (b_0 \Lam - 3k) (b_1 \Lam -3k)}{\sqrt{k} N_0^6 \Lam^2}
\bigl\{3(b_0 -b_1)^2 
\notag \\
&&
+ 12 kN_0^2 - 2(b_0+b_1) N_0^2 \Lam \bigr\}
\biggl(
\tan^{-1} 
\frac{3(b_0-b_1) + N_0^2\Lam}{6\sqrt{k} N_0}
+\tan^{-1}
\frac{3(b_1-b_0) + N_0^2\Lam}{6\sqrt{k} N_0}
\biggr) \biggr]\, .
\eea
On writing $N_c-N_\sg=n e^{i \ta}$, where $\ta$ is the angle Lefschetz thimble make 
with the real $N_c$-axis, then the above integration can be performed easily. 
It gives the following 
\bea
\label{eq:GabTrans}
G[b_0,b_1] \approx 
\sum_\sg n_\sg 
\sqrt{\frac{3 i V_3}{4 N_\sg \lvert \left(S_1\right)_{N_cN_c} \rvert}}
\exp\left(i \ta_\sg + \frac{iS_1^{\rm on-shell}}{\hbar}\right)
\biggl[
1 - \frac{N_\sg}{18} \int_0^1 {\rm d}t A_2 (N_\sg)
+ {\cal O}(\al^2) 
\biggr] \, ,
\eea
where $N_\sg$ and to first order in $\al$ is given 
by eq. (\ref{eq:N1_final}), $S_1^{\rm on-shell}$ is given in eq. (\ref{eq:S1_onshell}), 
$A_2(N_\sg)$ is given in eq. (\ref{eq:IntA2t}) and 
$\left(S_1\right)_{N_cN_c}$ is given in eq. (\ref{eq:S0NNsadpt}). 
This is a general expression for the transitional amplitude 
and valid for various kind of boundary conditions in the saddle 
point approximation to first order in $\al$. 
The corrections coming from novel Gauss-Bonnet gravity 
are present in $S_1^{\rm on-shell}$, $N_\sg$, 
$A_2(N_\sg)$ and $\left(S_1\right)_{N_cN_c}$.

%%%%%%%%%%%%%%%%%%%%%%%%%%%%%%%%%%%%%%%%%%%%%%%
\section{Classical Universe}
\label{cuni}
%%%%%%%%%%%%%%%%%%%%%%%%%%%%%%%%%%%%%%%%%%%%%%%

Here we study the transition probabilities in the classical Universe. This usually happens 
when $b_1>b_0 > (3k/\Lam)$. In this scenario the zeroth order saddle point solution 
for $N_c$ given in eq. (\ref{eq:Nc0_soln}) indicate the saddles are real. This furthermore 
leads to real on-shell zeroth order action as can be seen from eq. (\ref{eq:S0act_onshell}).
For each of the saddle point one can compute the second variation of action 
with respect to $N_c$ to find the directions of the steepest ascent and descent flows. 
In the case of classical boundary conditions the second variation at saddle point is real. 
This is easy to see from eq. (\ref{eq:S0NNsadpt}). From eq. (\ref{eq:changeH})
it immediately implies that $\rho=0$. This then translates into expression for 
$\ta$. 
\beq
\label{eq:StDeAe}
\ta_{\rm des} = k \pi + \frac{\pi}{4} \, ,
\hspace{5mm}
\ta_{\rm aes} = k \pi - \frac{\pi}{4} \, .
\eeq
The steepest descent/ascent flow-lines have a nice property that 
the imaginary part of $I$ (which is $H$) remains constant along them. This feature 
can be exploited to find them. We use this property to plot these flow lines
in the complex $N_c$-plane. For purpose of better understanding the things
we considered the following values of parameters: 
$(8\pi G) =1$, $k=1$, $\Lam=3$. 
The novel-Gauss-Bonnet parameter has mass-dimensions $M^{-2}$. 
If in eq. (\ref{eq:act}) we take $G$ inside bracket then we see 
$\al/G$ is dimensionless. This allow us to write 
$\al = \tilde{\al}/M_P^2$, where $M_P$ is Planck mass, and 
$\tilde{\al}$ is dimensionless. As in our convention $(8\pi G) =1$, 
this means $M_P^2 = 8\pi$. This then implies 
$\al = \tilde{\al}/(8\pi)$. 
%
%%%%%%%%%%%%%%%%%%%%%%%%%%%%%%%%%%%%%%%%%%%%%%%%%%%%%%%
\begin{figure}[h]
\centerline{
\vspace{0pt}
\centering
\includegraphics[width=5in]{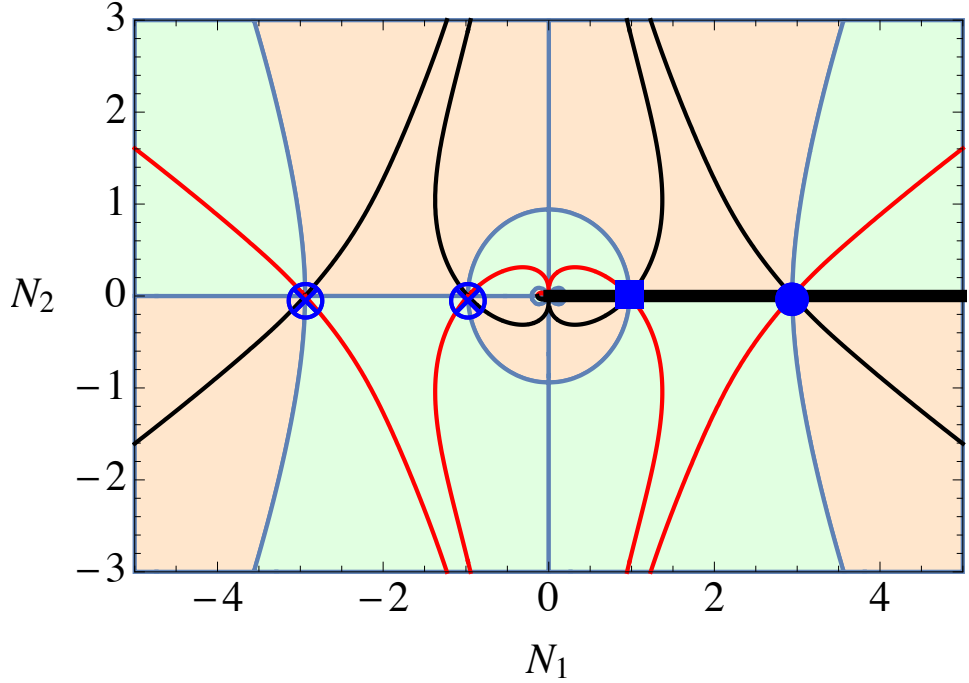}
}
 \caption[]{
In this plot we consider the case of classical boundary conditions:
$b_0=2$ and $b_1=5$. We take parameter values to be 
$(8\pi G) =1$, $k=1$, $\Lam=3$, while $\al=10^{-1} (8\pi)^{-1}$.
Here the red lines are steepest descent lines (thimbles ${\cal J}_\sg$), while thin black lines 
are steepest ascent lines denoted by ${\cal K}_\sg$. The saddle points $N_\sg$ are shown by 
blue. The two blue cross-circle are irrelevant saddle points, while the relevant 
saddle points are shown with blue-square and blue-dot respectively. 
Along the red and black lines $H$ remains 
constant and is equal to the value of $H(N_\sg)$. The light-green region has 
$h<h(N_\sg)$, while the light-orange region has $h>h(N_\sg)$. The 
boundary of these region is depicted by light-blue lines 
and along them $h = h(N_\sg)$. The original contour of 
integration $(0,\infty^+)$ is shown by thick black line. 
}
\label{fig:CUnGB}
\end{figure}
%%%%%%%%%%%%%%%%%%%%%%%%%%%%%%%%%%%%%%%%%%%%%%%%%%%%%%%

In the case of classical boundary conditions  $b_1>b_0 > (3k/\Lam)$, 
all the four saddle points lie on real axis: two are positive while two are negative. 
The two positive ones lie on the original integration contour $(0,\infty^+)$
and become relevant saddle point. The steepest descent paths passing through 
them will be relevant thimbles which both will contribute in the 
Lorentzian path integral. In the case of novel-GB gravity we notice that 
in the first order perturbation the saddle points have shifted compared to their
position in case of pure Einstein-Hilbert gravity \cite{Feldbrugge:2017kzv}.
We considered a simple example to study this situation where we depict 
the steepest descent/ascent flow lines (red/black lines), saddle points 
(blue cross-circle, blue-square, blue-dot), 
forbidden/allowed region (light-orange/light-green region) in figure \ref{fig:CUnGB}. 
The light-green region has $h < h(N_\sg)$ while light-orange region has $h >h(N_\sg)$.

The first relevant saddle starts from origin, circles around a bit in first quadrant, 
passes through blue-square then asymptotes to negative imaginary axis. 
The second relevant thimbles runs up from negative imaginary axis, passes 
through blue-dot and asymptotes to infinity at an angle $\pi/6$. 
Both these thimbles contribute to the Lorentizian path-integral and 
their sum is deformable to the original contour of integration
as explained in subsection \ref{choice} and in \cite{Feldbrugge:2017kzv}. 
The integral being absolutely convergent 
along the steepest descent lines naturally leads to a generalization of 
Wick rotation and correct answer for the Lorentzian path-integral.

For the case of classical boundary conditions, the relevant saddles 
and their corresponding steepest descent flow lines will have 
\beq
\label{eq:relevTH}
n_{\color{blue}{\small\CIRCLE}, \color{blue}{\blacksquare}} = 1 \, ,
\hspace{5mm}
\ta_{\color{blue}{\blacksquare}} = - \frac{\pi}{4}\, ,
\hspace{5mm}
\ta_{\color{blue}{\small\CIRCLE}}= \frac{\pi}{4} \, .
\eeq
If we define shorthand variables (to avoid clutter) 
\beq
\label{eq:shortuv}
u = \sqrt{\frac{b_0 \Lam}{3k} -1} \, , 
\hspace{5mm}
v = \sqrt{\frac{b_1 \Lam}{3k} -1} \, ,
\eeq
then in terms of them one can express saddle points, on-shell action and 
second variation in a compact form. They are given by following
\bea
\label{eq:sad_compact1}
&&
N_{\color{blue}{\blacksquare}}
= \frac{3\sqrt{k}(v-u)}{\Lam} 
+ \frac{\al \sqrt{k}}{2} \left[
15(v-u) + \frac{v-u}{uv}
+ 8 \left(\tan^{-1} v - \tan^{-1} u\right)
\right] + {\cal O}(\al^2)\, , 
 \\
\label{eq:sad_compact2}
&&
N_{\color{blue}{\small\CIRCLE}}
= \frac{3\sqrt{k}(v-u)}{\Lam} 
+ \frac{\al \sqrt{k}}{2} \left[
15(v-u) + \frac{v-u}{uv}
+ 8 \left(\tan^{-1} v - \tan^{-1} u\right)
\right] + {\cal O}(\al^2) \, ,
\\
\label{eq:ONS_S1}
&&
\left. S_1\right|_{\color{blue}{\blacksquare}}
= \frac{3 \pi ^2 \left(u^3-v^3\right)}{2 \Lambda }-\frac{1}{4} \pi ^2 \alpha  \left(5 u^3+3 u-5 v^3-3 v\right)
+  {\cal O}(\al^2) \, ,
\\
\label{eq:ONS_S2}
&&
\left. S_1\right|_{\color{blue}{\small\CIRCLE}}
= -\frac{3 \pi ^2 \left(u^3+v^3\right)}{2 \Lambda }+ \frac{1}{4} \pi ^2 \alpha  \left(5 u^3+3 u+5 v^3+3 v\right)
+ {\cal O}(\al^2) \, .
\eea
Using these one can write the leading order term for the transition amplitude 
in $1/\hbar$ expansion. This is given by,
\bea
\label{eq:transAmp}
G[b_0,b_1] = \frac{e^{i \pi/4}}{\sqrt{k uv}}
&&
\exp \biggl[
\frac{i \pi^2 \left\{-6 v^3 + \al \Lam v (3+5v^2) \right\}}{4\Lam \hbar}
\biggr]
\notag \\ 
&&
\times 
\cos \biggl[
\frac{\pi^2}{4\Lam \hbar} \left\{6 u^3 - \al \Lam (3u+5u^3) \right\} - \frac{\pi}{4}
\biggr] + {\cal O}(\al)\, ,
\eea
where we agree with the known results in $\al\to0$ results computed in 
\cite{Feldbrugge:2017kzv}. This is the leading term in $1/\hbar$ and first order
in $\al$. For the next order term, we will mention it in the Appendix \ref{gabbar}
due to its length.

%%%%%%%%%%%%%%%%%%%%%%%%%%%%%%%%%%%%%%%%%%%%%%%
\section{Summary and Conclusion}
\label{conc}
%%%%%%%%%%%%%%%%%%%%%%%%%%%%%%%%%%%%%%%%%%%%%%%

In this paper we study novel-Gauss-Bonnet (nGB) action by performing $D\to4$ limit 
carefully. We study this scenario in cosmology and consider a generalised FLRW Universe 
respecting homogeneity and isotropicity in arbitrary spacetime dimensions. We compute 
an action for scale-factor $a(t_p)$ and lapse $N_p(t_p)$ in the nGB gravity, where we notice 
that an integration by parts allow us to take the $D\to4$ limit smoothly without encountering 
divergences. The residual finite action obtained is used to study the classical and quantum 
aspects of theory in empty Universe. 

In the first part of paper we reproduce the results obtained in the paper \cite{Narain:2020qhh}
for classical cosmic evolution but for nonzero $k$. As in \cite{Narain:2020qhh}
we do a redefinition of scale factor $a$ and lapse $N_p$, thereby writing the theory 
in term of $q(t)$ and $N(t)$. The resulting action is a function of 
$N$, $q$ and $\dot{q}$ only, and doesn't contain any $t$-derivative of lapse $N$. 
Varying this action with respect to $q$ gives equation of motion for $q(t)$, while 
varying with respect to $N$ gives a constraint. We solve the equation of motion
for $q(t)$ perturbatively to first order in $\al$ for non-zero $k$ for given boundary 
conditions. On plugging this back in to action of theory, gives us an action for 
lapse $N$, which can be varied to obtain saddle points for $N$. This has to be done 
order by order. 

In the second part of paper we study the quantum aspects of the mini-superspace 
action of theory in the nGB gravity. We ask a straight-forward question 
what is the amplitude of transition from one $3$-geometry to another in the case 
when gravity is getting modified due to novel-Gauss-Bonnet term?
To answer this we study the path-integral of the mini-superspace theory 
by doing path-integration over $q(t)$ and lapse $N$. We study this 
directly in Lorentzian signature without doing 
a Wick-rotation of time co-ordinate to obtain Euclideanised theory. 
This is Lorentzian quantum cosmology of novel-GB gravity. 
We follow the strategy described in \cite{Halliwell:1988ik,Halliwell:1989dy,Feldbrugge:2017kzv}
to analyse the path-integral in the mini-superspace approximation.
We study this in gauge $\dot{N}=0$ (implying $N(t)=N_c$, a $t$-independent parameter).
This path-integral consist of two segments: path-integral over $q(t)$ 
and an ordinary integral over $N_c$. We study the former using WKB 
approximation while for the later we use combination of Picard-Lefschetz methods
and WKB to compute the transition amplitude. 
We follow the footsteps of formalism developed 
in \cite{Narain:2019qcj} to compute this transition probability.

Then do the path-integral for $q(t)$ using WKB we write $q(t) = q_b (t) + Q(t)$,
where $Q(t)$ is fluctuation around the background solution $q_b(t)$ which is 
computed perturbatively to first order in $\al$. This gives us a 
path-integral over $Q(t)$ satisfying vanishing boundary conditions.
In sub-section \ref{Qint} we perform the $Q$-integral. Due to 
non-linear nature of the original mini-superspace action our 
abilities are limited and we compute this $Q$-integral perturbatively 
to first order in $\al$, following the strategy outlined in \cite{Narain:2019qcj}.

For the $N_c$ integral we make use of techniques of Picard-Lefschetz methods
to analyse the integral in complex $N_c$ plane. For a generic set of boundary 
conditions we compute the saddle points in complex $N_c$ plane. We make 
use of flow equations the find the behavior of Morse function $h$ and $H$. 
It is noticed that $H$ remains constant along the steepest descent/ascent flow 
lines, a property which is used later to numerically draw a graph on the complex 
$N_c$ plane. Depending on boundary conditions it is seen that not all saddle points 
are relevant, as steepest ascent paths from only some will interest the original 
integration contour. The steepest descent paths from these relevant saddles 
will constitute the relevant thimbles contributing to the path-integral. The original 
contour can be deformed in to a contour passing through these relevant thimbles.
The $N_c$-integral is then performed along these thimbles, taking contribution 
from all relevant thimbles. This is a generalization of Wick rotation. 
We obtain an expression for transition amplitude $G[b_0, b_1]$ to first 
order in $\al$ and in $1/\hbar$ expansion. This is given in 
eq. (\ref{eq:GabTrans}). 

We use this to investigate the case of classical boundary conditions 
where $b_1>b_0 > (3k/\Lam)$. In this case the saddle points are all real 
and their corresponding on-shell action is also real. In this case we compute 
numerically the flow lines, and determine the angles they make with real axis. 
Out of the four real saddles (two positive and two negative), only the two
positive ones are relevant, as they lie on the original integration contour. 
This implies that only two steepest descent curves are relevant, which will 
contribute in the $N_c$-integral. Combining all the ingredients we were 
finally able to write the leading order term in the transition amplitude for the
case of classical boundary conditions, which is given in eq. (\ref{eq:transAmp}). 
In the limit $\al\to0$ this agrees with the result in \cite{Feldbrugge:2017kzv},
and we write the next order terms in the appendix. We notice that 
novel-GB gravity gives non-trivial correction to transition amplitude 
even though our analysis was done perturbatively.

%%%%%%%%%%%%%%%%%%%%%%%%%%%%%%%%%%%%%%%%%%%%%%%
\bigskip
\centerline{\bf Acknowledgements} 
%%%%%%%%%%%%%%%%%%%%%%%%%%%%%%%%%%%%%%%%%%%%%%%

GN will like to thank Nirmalya Kajuri and Avinash Raju for discussion.
GN is supported by ``Zhuoyue" (distinguished) Fellowship (ZYBH2018-03).
H. Q. Z. is supported by the National Natural Science Foundation of China (Grants
No. 11675140, No. 11705005, and No. 11875095).

%%%%%%%%%%%%%%%%%%%%%%%%%%%%%%%%%%%%%%%%%%%%%%%
\appendix
%%%%%%%%%%%%%%%%%%%%%%%%%%%%%%%%%%%%%%%%%%%%%%%

%%%%%%%%%%%%%%%%%%%%%%%%%%%%%%%%%%%%%%%%%%%%%%%
\section{$S_{N_cN_c}$ at saddles, $A_2$ and ${\cal O}(\al)$ terms}
\label{gabbar}
%%%%%%%%%%%%%%%%%%%%%%%%%%%%%%%%%%%%%%%%%%%%%%%

Here we write the expression for second variation of action at saddle points.
For the classical boundary conditions the second variation at the two relevant saddle points 
is given by,
\bea
\label{eq:S2varNN}
&&
\left. \left(S_1\right)_{N_c N_c} \right|_{\color{blue}{\blacksquare}}
= - \frac{2 \sqrt{k} \Lam uv}{v-u} 
+ \frac{\sqrt{k} \Lam^2 \al}{3(v-u)} \biggl[
\frac{v^2 + uv + u^2}{uv} -9 uv
+ \frac{24 u^2 v^2 \left(\tan^{-1} v - \tan^{-1} u\right)}{(v-u)}
\biggr] \, ,
\notag\\
&&
\left. \left(S_1\right)_{N_c N_c} \right|_{\color{blue}{\small\CIRCLE}}
= - \frac{2 \sqrt{k} \Lam uv}{v+u} 
+ \frac{\sqrt{k} \Lam^2 \al}{3(v+u)} \biggl[
\frac{v^2 - uv + u^2}{uv} -9 uv
- \frac{24 u^2 v^2 \left(\tan^{-1} v + \tan^{-1} u\right)}{(v+u)}
\biggr]
\eea
In the computation of transition amplitude to first order in $\al$ 
we require to compute $A_2$ at the saddle points. Its expression 
at the two relevant saddle point it given by,
\bea
\label{eq:A2form1}
&&
\left. A_2 \right|_{\color{blue}{\blacksquare}}
= -\frac{5 \alpha  \Lambda ^3 (u+v) (u v+1)}{6 k^{3/2} \left(u^2+1\right)^2 \left(v^2+1\right)^2}
+\frac{\alpha  \Lambda ^4}{11664 k^{5/2} \left(u^2+1\right)^2 \left(v^2+1\right)^2 (u-v)^3}
\notag\\
&& 
\times
\biggl[\sqrt{k} (u-v) \left(u^4+2 u^2+v^4+2 v^2+2\right)+3 \left(u^2+v^2+2\right) \left(u^4+u^2 \left(4
v^2+6\right)+v^4+6 v^2+6\right)\biggr]
\notag \\
&&
+\frac{\alpha \Lambda ^2 (u v+1)}
{4 \sqrt{k} \left(u^2+1\right)^2 \left(v^2+1\right)^2(u-v)}
\biggl[u^4 \left(5 v^2+9\right)-10 u^3 \left(v^3+v\right)+u^2 \left(5 \left(v^2+2\right) v^2+13\right)
\notag \\
&&
-10 u \left(v^3+v\right)+9 v^4+13 v^2+8 \biggr]
+ \frac{13 \alpha  \Lambda ^2 \left(\tan ^{-1}u -\tan ^{-1}v\right)}{8 \sqrt{k}} \, ,
\\
\label{eq:A2form2}
&&
\left. A_2 \right|_{\color{blue}{\small\CIRCLE}}
= -\frac{5 \alpha  \Lambda ^3 (u-v) (u v-1)}{6 k^{3/2} \left(u^2+1\right)^2 \left(v^2+1\right)^2}
+\frac{\alpha  \Lambda ^4}{11664 k^{5/2} \left(u^2+1\right)^2 \left(v^2+1\right)^2 (u+v)^3}
\notag \\
&&
\times
\biggl[\sqrt{k} (u+v) \left(u^4+2 u^2+v^4+2 v^2+2\right)-3 \left(u^2+v^2+2\right) 
\left(u^4+u^2 \left(4v^2+6\right)+v^4+6 v^2+6\right)\biggr]
\notag \\
&&
+\frac{\alpha \Lambda ^2 (u v-1)}{4 \sqrt{k} \left(u^2+1\right)^2 \left(v^2+1\right)^2(u+v)}
\biggl[u^4 \left(5 v^2+9\right)+10 u^3 \left(v^3+v\right)+u^2 \left(5 \left(v^2+2\right)
v^2+13\right)
\notag \\
&&
+10 u \left(v^3+v\right)+9 v^4+13 v^2+8\biggr]
-\frac{13 \alpha  \Lambda ^2 \left(\tan ^{-1}u+\tan ^{-1}v\right)}{8 \sqrt{k}} \, .
\eea
The order $\al$ correction piece to the transition amplitude is given by,
\bea
\label{eq:NeOd_G}
&&
\left. G[b_0,b_1] \right|_{{\cal O}(\al)} 
= \frac{\alpha  \sqrt{u v}}{139968 u^3 v^3 \left(u^2+1\right)^2 \left(v^2+1\right)^2 }
\biggl[
\frac{\Lambda}{(u-v)^2} 
\notag \\
&&
\times
\biggl\{-9720 \Lambda  u^2 v^2 (u+v) (u v+1) (u-v)^3+\Lambda ^2 u^2 v^2 \bigl((u-v) 
\left(u^4+2 u^2+v^4+2v^2+2\right)
\notag \\
&&
+3 \left(u^2+v^2+2\right) \left(u^4+u^2 \left(4 v^2+6\right)+v^4+6 v^2+6\right)\bigr)
+2916 \bigl(u^7 v^3
\left(5 v^2+9\right)
\notag \\
&&
+u^6 \left(2 v^6+21 v^4+25 v^2+2\right)
+u^5 v^3 \left(5 v^4+3\right)+u^4 \left(21 v^6+56 v^4+47v^2+4\right)
\notag \\
&&
+u^3 v^3 \left(9 v^4+3 v^2-2\right)+u^2 \left(25 v^6+47 v^4+28 v^2+2\right)+2 \left(v^3+v\right)^2\bigr)
(u-v)^2\biggr\} 
\notag \\
&&
\times 
\exp \biggl\{-\frac{i \pi ^2}{4 \hbar \Lambda } 
\left(
6 v^3 - 6 u^3 + \al \Lam \left(3u -3v + 5u^3 - 5 v^3\right)
\right)\biggr\}
\notag \\
&&
- \frac{i \Lambda}{(u+v)^2} \biggl\{-9720 \Lambda  u^2 v^2
(u-v) (u v-1) (u+v)^3+\Lambda ^2 u^2 v^2 \bigl((u+v) \left(u^4+2 u^2+v^4+2 v^2+2\right)
\notag \\
&&
-3 \left(u^2+v^2+2\right)
\left(u^4+u^2 \left(4 v^2+6\right)+v^4+6 v^2+6\right)\bigr)
+2916 \bigl(u^7 v^3 \left(5 v^2+9\right)
\notag \\
&&
-u^6 \left(2v^6+21 v^4+25 v^2+2\right)+u^5 v^3 \left(5 v^4+3\right)-u^4 \left(21 v^6+56 v^4+47 v^2+4\right)
\notag \\
&&
+u^3 v^3 \left(9 v^4+3v^2-2\right)-u^2 \left(25 v^6+47 v^4+28 v^2+2\right)
-2 \left(v^3+v\right)^2\bigr) (u+v)^2\biggr\} 
\notag \\
&&
\times
\exp \biggl\{ \frac{i\pi ^2}{4\hbar\Lambda } 
\left(
- 6u^3 -6 v^3 + \al \Lam(3u + 3v + 5u^3 + 5v^3)
\right)\biggr\}
\biggr]
\notag \\
&&
+ \frac{i \alpha  \Lambda}{96 \sqrt{u v}} 
\biggl[\frac{\left(13 u^2-70 u v+13 v^2-32\right)}{u+v}\tan ^{-1}(u) 
\notag \\
&&
\times 
\exp \biggl\{\frac{i \pi ^2}{4\hbar \Lambda } 
\left(
-6v^3 - 6u^3 + \al \Lam (3u + 3v + 5u^3 + 5v^3)
\right)\biggr\}
\notag \\
&&
-i \frac{\left(13 u^2+70 u v+13 v^2-32\right)}{u-v} \tan ^{-1}(u) 
\exp \biggl\{-\frac{i \pi ^2}{4 \hbar \Lambda }
\left(
6v^3 - 6 u^3 + \al \Lam(3u - 3v + 5u^3 - 5 v^3)\right)\biggr\}
\notag \\
&&
+\frac{\left(13 u^2-70 u v+13 v^2-32\right)}{u+v}\tan ^{-1}(v) 
\exp \biggl\{\frac{i \pi ^2}{4 \hbar \Lambda }
\left(
- 6u^3 -6 v^3 + \al \Lam(3u + 3v + 5u^3 + 5v^3)\right)\biggr\}
\notag \\
&&
+i \frac{\left(13 u^2+70 u v+13 v^2-32\right)}{u-v} \tan ^{-1}(v) 
\exp \biggl\{-\frac{i \pi ^2}{4 \hbar \Lambda }
\left(
6v^3 - 6 u^3 + \al \Lam(3u - 3v + 5u^3 - 5 v^3)\right)\biggr\}\biggr] \, .
\eea

%%%%%%%%%%%%%%%%%%%%%%%%%%%%%%%%%%%%%%%%%%%%%%%

\end{document}